\documentclass[a4paper,11pt]{article}
\usepackage{jheppub} 
\usepackage{comment}
\usepackage{tikz}
\usepackage{physics}
\usepackage{slashed, xcolor, color}
\usepackage{amsmath}
\usepackage[compat=1.0.0]{tikz-feynman}
\usepackage{float}
\usepackage{multirow}
\usepackage{nicematrix}
\usepackage{dsfont}
\usepackage[T1]{fontenc}

\usepackage{cleveref}
\crefname{table}{Table}{Tables}
\crefname{equation}{Eq.}{Eqs.}
\crefname{appendix}{App.}{Apps.}
\crefname{section}{Sec.}{Secs.}
\crefname{figure}{Fig.}{Figs.}

\newcommand{\dneff}{\Delta N_{\mathrm{eff}}}
\newcommand{\lc}{\mathcal{L}}

\title{Baryogenesis through Asymmetric Reheating in the Mirror Twin Higgs}

\author{Gonzalo Alonso-\'Alvarez,$^{1,2}$}
\author{David Curtin,$^1$}
\author{Andrija Rasovic$^1$ and}
\author{Zhihan Yuan$^1$}

\affiliation{$^1$Department of Physics, University of Toronto, Toronto, ON M5S 1A7, Canada}
\affiliation{$^2$McGill University Department of Physics, 3600 Rue University, Montr\'eal, QC H3A 2T8, Canada}

\emailAdd{gonzalo.alonso@utoronto.ca}
\emailAdd{dcurtin@physics.utoronto.ca}
\emailAdd{a.rasovic@mail.utoronto.ca}
\emailAdd{ly.yuan@mail.utoronto.ca}

\abstract{ 
We present the $\nu\phi$MTH, a Mirror Twin Higgs (MTH) model 
realizing asymmetric reheating, baryogenesis and twin-baryogenesis through the out-of-equilibrium decay of a right-handed neutrino without any hard $\mathbb{Z}_2$ breaking.
The MTH is the simplest Neutral Naturalness solution to the little hierarchy problem and predicts the existence of a twin dark sector related to the Standard Model (SM) by a $\mathbb{Z}_2$ symmetry that is only softly broken by a higher twin Higgs vacuum expectation value.  
The asymmetric reheating cools the twin sector compared to the visible one, thus evading cosmological bounds on $\dneff$.
The addition of (twin-)colored scalars allows for the generation of the visible baryon asymmetry and, by the virtue of the $\mathbb{Z}_2$ symmetry, also results in the generation of a twin baryon asymmetry.
We identify a unique scenario with top-philic couplings for the new scalars that can satisfy all cosmological, proton decay and LHC constraints; yield the observed SM baryon asymmetry; and generate a wide range of possible twin baryon DM fractions, from negligible to unity. The viable regime of the theory contains several hints as to the possible structure of the Twin Higgs UV completion. 
Our results motivate the search for the rich cosmological and astrophysical signatures of twin baryons, and atomic dark matter more generally, at cosmological, galactic and stellar scales.
}

\begin{document}
\maketitle
\flushbottom

\section{Introduction}
\label{sec:intro}

Despite its striking success, the Standard Model (SM) remains an incomplete description of fundamental physics. While its quantitative predictions are mostly fulfilled to impressive accuracy, it completely fails to account for a number of observed phenomena, including the large hierarchy between the electroweak scale and the Planck scale, the matter-antimatter asymmetry in the universe, dark matter (DM), and neutrino masses. There must therefore be physics beyond the Standard Model (BSM).

The Higgs boson mass (and hence its vacuum expectation value) is quadratically sensitive to the scale of BSM physics. The naive expectation within the SM is therefore an electroweak scale far above the observed value, which is referred to as the Hierarchy Problem. 
Canonical solutions to the hierarchy problem include variations of supersymmetry \cite{Martin:1997ns}, Composite Higgs \cite{Csaki:2018muy} and Little Higgs models \cite{Arkani-Hamed:2002iiv,Schmaltz:2004de}. 
A general feature of all these models is that the symmetry which protects the Higgs mass predicts the existence of new states related to SM particles. 
While details vary, the partner of the top quark, with its large Higgs coupling, is required to have a mass at or below the TeV scale 
to avoid excessive fine tuning of the model parameters.
Since these partners have the same quantum numbers as the corresponding SM particle, and since the Large Hadron Collider (LHC) has excellent reach for discovering new QCD-charged states, null results from searches to date \cite{ATLAS:2017drc,ATLAS:2022ihe,CMS:2017jrd,CMS:2017kil} imply a higher mass scale for these top partners, if they exist.
In the absence of an additional mechanism that addresses this issue, the model parameters have to be fine-tuned, leading to the so-called ``little hierarchy problem".

A different approach to the (little) hierarchy problem is Neutral Naturalness~\cite{Chacko:2005pe, Burdman:2006tz, Cai:2008au, Craig:2014aea, Cohen:2018mgv, Cheng:2018gvu}, see also~\cite{Batell:2022pzc, Craig:2023lyt} for recent reviews. This refers to a class of models where the symmetry partners are neutral under SM color (and commonly under all SM gauge groups), hence evading the stringent LHC constraints on colored top partners. The canonical representative of this class of models is the Mirror Twin Higgs (MTH) scenario~\cite{Chacko:2005pe, Chacko:2016hvu, Craig:2016lyx}. In the MTH framework, a hidden twin sector is related to the SM via a discrete $\mathbb{Z}_2$ symmetry. This hidden sector contains copies of all SM fields, which carry no SM charges but are instead charged under their own copies of the SM gauge groups. 
The Higgs mass is protected by an approximate accidental $SU(4)$ symmetry that arises in the $SU(2) \cross SU(2) \cross \mathbb{Z}_2$ symmetric Lagrangian at the quadratic level.
In the infrarred (IR), meaning at the TeV scale and below, the only required interaction between the two sectors is the Higgs portal which mixes the visible and twin Higgs bosons.
If the $\mathbb{Z}_2$ symmetry remains unbroken, the 125 GeV light pNGB Higgs would be an equal admixture of scalars in both sectors.
A soft $\mathbb{Z}_2$ breaking Higgs mass 
ensures that the light Higgs 
lives mostly in the visible sector with an invisible admixture $\sim v^2/f^2$, satisfying Higgs coupling constraints~\cite{CMS:2018uag, ATLAS:2022vkf}. The twin Higgs vacuum expectation value (vev) $f$ is a few times greater than the visible Higgs vev $v$, $f/v \gtrsim 3$, and the minimal scenario carries a modest tuning $\sim 2 v^2/f^2$, which can be significantly ameliorated in extended models~\cite{Beauchesne:2015lva, Harnik:2016koz, Csaki:2019qgb, Durieux:2022sgm}.

The original MTH model is however excluded by cosmology: the light twin photons and neutrinos result in a $\dneff$ value of around 6, much higher than the bound of $\dneff \lesssim  0.3$ from Big Bang Nucleosynthesis (BBN) and Cosmic Microwave Background (CMB) measurements \cite{Planck:2018vyg}. 
Broadly, this can be addressed in two ways. One possible solution is to introduce extra hard $\mathbb{Z}_2$ breaking into the low-energy effective Twin Higgs theory. This can involve removing the light twin degrees of freedom completely, as in the Fraternal Twin Higgs~\cite{Craig:2015pha}, or increasing their Yukawa couplings relative to the SM values~\cite{Barbieri:2016zxn, Barbieri:2017opf}.
Long-lived particle signatures at the LHC due to twin hadron decays to the visible sector are a generic prediction of this approach. 
On the other hand, additional sources of $\mathbb{Z}_2$ breaking can be avoided by invoking a cosmological solution, leading to the \emph{asymmetrically reheated MTH}~\cite{Chacko:2016hvu, Craig:2016lyx, Ireland:2022quc}.
In this setup, additional heavy particles (`reheatons') freeze out, lead to an early period of matter domination, and then decay out of equilibrium. 
Due to the mass differences between the two sectors generated by the existing soft $\mathbb{Z}_2$ breaking, this decay preferentially reheats the visible sector.
The relatively colder twin sector can thus generate a viably small $\dneff$.

The asymmetrically reheated MTH is an attractive scenario. The $\mathbb{Z}_2$ symmetry that protects the Higgs mass is fully preserved, except for the \emph{minimal} soft breaking in the scalar sector that is required to reproduce the broadly SM-like Higgs boson that we observe.\footnote{This is in perfect analogy to the necessity of soft SUSY breaking terms in the MSSM. Furthermore, just like in SUSY, a variety of mechanisms can generate these breaking terms dynamically~\cite{Beauchesne:2015lva, Harnik:2016koz, Yu:2016bku, Yu:2016swa}.}
The absence of hard $\mathbb{Z}_2$ breakings also allow for simpler UV completions of the model above 5-10 TeV (see 
\cite{Falkowski:2006qq,Chang:2006ra, Craig:2013fga,  Katz:2016wtw, Badziak:2017syq, Badziak:2017kjk, Badziak:2017wxn, Asadi:2018abu, Choi:2023eus}
and
\cite{Geller:2014kta, Barbieri:2015lqa, Low:2015nqa, Xu:2018ofw, Dillon:2018wye, Xu:2019xuo}
for supersymmetric and composite constructions respectively).
Therefore, we 
are led to this cosmologically non-minimal solution
to maximally preserve the discrete symmetry which is at the heart of the twin mechanism.
This, in turn, invites us to seriously consider all the consequences that this untarnished $\mathbb{Z}_2$ symmetry would have for our universe.\footnote{Another attractive approach is to consider the $\mathbb{Z}_2$ symmetry together with the possible solution to the SM \& twin Flavor Problem. The larger Yukawas for the non-top twin fermions can arise as a result of (possibly soft) $\mathbb{Z}_2$ breaking in some flavor sector, which can radiatively generate the necessary $\mathbb{Z}_2$-breaking in the Higgs sector while naturally preserving the twin protection mechanism for top contributions only~\cite{Barbieri:2016zxn, Barbieri:2017opf}.} In this work, we focus on the generation of the matter-antimatter asymmetry of the universe.

The presence of baryons but no antibaryons in our universe requires an explanation beyond the SM.
The out-of-equilibrium decay of a heavy parent particle in the early universe is a very common way of generating a baryon (or other) asymmetry, see e.g.~\cite{Davidson:2008bu,Cui:2012jh, Cui:2014twa}. Since the $\mathbb{Z}_2$-symmetric MTH already requires an out-of-equilibrium decay to realize asymmetric reheating, one is prompted to ask whether this same decay could also be responsible for generating the SM baryon asymmetry. 
A natural expectation is that this would also generate a twin baryon asymmetry, which in the $\mathbb{Z}_2$-symmetric MTH would constitute an atomic dark matter (aDM)~\cite{Kaplan:2009de} sub-component of DM. This is 
largely unconstrained by existing self-interaction constraints for $\hat r \equiv \Omega_\mathrm{aDM}/\Omega_\mathrm{DM} \lesssim \mathcal{O}(0.2)$~\cite{Randall:2008ppe}
but generates a plethora of cosmological~\cite{
Cyr-Racine:2013fsa,
Bansal:2022qbi,
Bansal:2021dfh,
Zu:2023rmc} and complex astrophysical~\cite{
Fan:2013yva,Fan:2013tia,McCullough:2013jma,Randall:2014kta,Schutz:2017tfp, Buch:2018qdr,Ghalsasi:2017jna, Ryan:2021dis, Gurian:2021qhk, Ryan:2021tgw, Foot:2013lxa,  Foot:2014uba, Foot:2015mqa,Chashchina:2016wle, Foot:2017dgx, Foot:2018dhy, Foot:2016wvj, Foot:2013vna, Roy:2023zar, Gemmell:2023trd, Curtin:2019ngc, Curtin:2019lhm, Howe:2021neq, Winch:2020cju, Hippert:2021fch, Hippert:2022snq, Curtin:2020tkm, Ryan:2022hku, Pollack:2014rja, Shandera:2018xkn, Singh:2020wiq, Gurian:2022nbx, Fernandez:2022zmc, Bai:2023mfi} signatures in our universe which depend sensitively on $\hat r$.
Understanding the detailed properties of an MTH theory which realizes this scenario, and obtaining predictions or theoretically favored ranges for the twin baryon DM fraction, is therefore highly motivated.

Our approach to studying baryogenesis in the Twin Higgs should be contrasted to previous investigations. 
Several studies implement hard $\mathbb{Z}_2$ breakings and generate \emph{all} of the observed DM density from a twin baryon asymmetry generated in an out-of-equilibrium decay together with SM baryons, which requires various departures from the minimal MTH spectrum to accommodate DM self-interaction constraints, account for the full DM relic density, and eliminate dark radiation~\cite{Farina:2015uea, Farina:2016ndq,  Earl:2019wjw, Feng:2020urb, Kilic:2021zqu}.
Utilizing phase transitions in a variety of electroweak baryogenesis, rather than particle decays, has also been investigated~\cite{Fujikura:2018duw, Badziak:2022ltm}, but the Minimal Twin Higgs setup seems unsuitable to generate the visible baryon asymmetry in this way, and twin baryons are only generated if the twin phase transition is affected by components of the UV completion or modified with hard $\mathbb{Z}_2$ breakings. 
The closest analogue to our approach in the past literature is a recent MTH model proposal~\cite{Beauchesne:2021opx}, where two successive out-of-equilibrium decays successively generate visible/twin baryon asymmetries, and then the asymmetric reheating, respectively. While this scenario shares some ingredients with our model, we will unify the asymmetric reheating and baryon number generation mechanisms in the same out-of-equilibrium process.

Our setup can be placed in this context by emphasizing that we aim to preserve the \emph{maximally $\mathbb{Z}_2$-symmetric} nature of the MTH, \emph{minimally} modifying the hence required  asymmetric reheating mechanism to generate the \emph{visible} baryon asymmetry and obtain \emph{predictions for the DM fraction of twin baryons} that generate rich but still poorly understood cosmological and astrophysical signatures, with the rest of DM being assumed to be made up of some unspecified, possibly symmetric WIMP-like candidate~\cite{Hochberg:2018vdo, Badziak:2019zys, Ahmed:2020hiw, Curtin:2021alk, Badziak:2022eag}.\footnote{
Furthermore, our model can be easily extended to make the $\mathbb{Z}_2$-breaking spontaneous, as e.g. in ~\cite{Beauchesne:2021opx}.}

The model that we propose and coin $\nu\phi$MTH is a minimal extension of the $\nu$MTH setup first proposed 
in~\cite{Chacko:2016hvu}. 
Just like in the $\nu$MTH, a twin-version of the type-1 seesaw mechanism generates visible and twin neutrino masses, resulting in right-handed neutrino (RHN) states that can decay to both visible and twin states.
The most long-lived $N$ decays out-of-equilibrium and reheats both sectors. This reheating is asymmetric purely due to mass differences in the final states generated by  the minimal soft $\mathbb{Z}_2$-breaking $f > v$. 
Our model features an additional $\gtrsim$~TeV-scale \mbox{(twin-)} colored \mbox{(twin-)} scalar $\phi$, coupling to \mbox{(twin-)} quarks as well as the RHNs, which allows a small fraction of $N$-decays to generate \mbox{(twin-)} baryon asymmetries. 
Depending on the model's parameters, this can generate all of the required DM in the form of twin baryons; twin-baryon DM fractions of $\hat r \sim 0.25$ as expected by $\mathbb{Z}_2$ symmetric number densities; lower values of $\hat r \lesssim \mathcal{O}(0.1 - 1\%)$ that are particularly interesting for astrophysical signatures while avoiding cosmological bounds; or tiny fractions that may be undetectable. 
Lower twin baryon number densities are again purely brought about by mass differences between the two sector particles that result from the minimal soft $\mathbb{Z}_2$-breaking.\footnote{During the completion of this work, Ref.~\cite{Bittar:2023kdl} appeared on the arXiv, which also considers baryogenesis in a modified MTH model that shares some similar ingredients with our $\nu\phi$MTH model. However, in that work, the mechanism of asymmetric reheating is left undetermined, with an exclusive focus on the generation of visible and twin baryon number. Furthermore, the means by which differences in the twin and visible baryon abundances are generated is a phase difference between the visible and twin sector couplings. At the time of writing, the origin of that phase difference is unexplained.}

This paper is structured as follows.
In \cref{sec:review} we briefly review the asymmetric reheating mechanism for the MTH, generate updated $\dneff$ constraints in the plane of the reheaton mass $m_N$ and the asymmetric reheating temperature $T_{A,R}$, exploring for the first time the higher-mass regime for the reheaton in the $\nu$MTH, and comment on the important relationship between twin baryogenesis from reheaton decay and twin BBN.
We introduce our $\nu\phi$MTH model in \cref{sec:2-gen-model}, and discuss LHC and proton decay constraints on the new scalar $\phi$ as well as the flavor alignment requirements and the modified cosmological history of this model.
In \cref{sec:baryogenesis} we carefully discuss the different decay scenarios for (twin-) baryon generation in the $\nu\phi$MTH, demonstrating how different twin baryon number densities are produced and the corresponding required parameters of the model.
We outline the lessons for Neutral Naturalness model building and conclude in \cref{sec:conclusion}.


\section{The Minimal Twin Higgs with Asymmetric Reheating}\label{sec:review}

In this section we briefly review the Mirror Twin Higgs (MTH) model~\cite{Chacko:2005pe}, its cosmology, and the implementation of the asymmetric reheating mechanism in the $\nu$MTH model~\cite{Chacko:2016hvu}. We then revisit the $\dneff$ predictions for this model in light of the most recent Planck bounds~\cite{Planck:2018vyg}, with a particular emphasis on the previously unexplored high-mass reheaton regime, and the role of twin BBN. We also review the properties of twin baryons in the MTH model. 

\subsection{The Twin Higgs Mechanism}

The MTH model, first proposed in \cite{Chacko:2005pe}, introduces a hidden twin sector related to the SM via a discrete $\mathbb{Z}_2$ symmetry, containing twin copies of all SM particles and gauge forces. 
We denote visible (twin) particles with subscript $A$ ($B$). 
Defining $H = (H_A, H_B)$, where $H_{A,B}$ are the visible and twin Higgs doublets, the scalar potential
\begin{gather}
V(H)=-m^2 H^{\dagger} H+\sigma\left(H^{\dagger} H\right)^2
\end{gather}
satisfies an approximate $SU(4)$ symmetry. 
In order to comply with experimental bounds on SM Higgs couplings, we introduce a small quartic term that breaks $SU(4) \to SU(2)_A \times SU(2)_B \times \mathbb{Z}_2$, and a soft $\mathbb{Z}_2$ breaking mass:
\begin{gather}
\begin{aligned}
V(H)=  -m^2|{H}|^2+\sigma|{H}|^4 
 +\mu^2\left(\left|H_A\right|^2-\left|H_B\right|^2\right)+\delta\left(\left|H_A\right|^4+\left|H_B\right|^4\right).
\end{aligned}
\end{gather}
The twin (visible) Higgs $H_B$ ($H_A$) acquires a vev $f$ ($v$), with the hierarchy $f/v \gtrsim 3$ requiring a choice of $\mu$ that is tuned at the level of $\sim 2 v^2/f^2$. 
The observed 125 GeV SM Higgs $h \approx H_A - (v/f) H_B + \mathcal{O}(v^2/f^2)$ (perturbed around their respective vevs) is identified as a pseudo-Nambu-Goldstone boson (pNGB) of the approximate $SU(4)$ symmetry, and its mass is thus protected from large corrections. To see this explicitly, consider the visible and twin top quark Yukawa couplings,
\begin{gather}
    \lc\supset y_{t,A} H_A \bar{Q}_A q_A  + y_{t,B} H_B \bar{Q}_B q_B \, ,  
\end{gather}
where $Q_B$ are left-handed twin quarks and $q_B$ are right-handed twin quarks. The $\mathbb{Z}_2$ symmetry ensures that $y_{t,A}=y_{t,B}=y_t$, so the quadratically divergent corrections to the Higgs potential from the top and twin top quarks take the form
\begin{gather}
    \Delta V=\frac{3 y_t^2}{8 \pi^2} \Lambda^2\left(H_A^{\dagger} H_A+H_B^{\dagger} H_B\right)=\frac{3 y_t^2}{8 \pi^2} \Lambda^2 H^{\dagger} H.
\end{gather}
Here, $\Lambda \sim 5 - 10~\mathrm{TeV}$ is the UV cutoff scale of the effective low-energy Twin Higgs theory. This correction respects the approximate global $SU(4)$ symmetry, and does not contribute to the mass of our pNGB, thereby solving the little hierarchy problem.
Since the two sectors only interact through the Higgs portal, the main collider signatures of the MTH are Higgs coupling deviations and invisible decays into the hidden sector as a result of the mixing between the two Higgs doublets. Both occur at the $\sim v^2/f^2$ order~\cite{Burdman:2014zta}.
We focus on $f \sim 3 - 7$ as the most natural scenarios in the minimal model with tuning $\sim 2 v^2/f^2$, but higher twin higgs vevs could be motivated in non-minimal models that naturally generate large vev hierarchies~\cite{Beauchesne:2015lva, Harnik:2016koz, Csaki:2019qgb, Durieux:2022sgm}.

\subsection{Asymmetric Reheating with the $\nu$MTH}

The original Twin Higgs model is excluded by cosmology. More specifically, the light degrees of freedom in the twin sector result in a $\dneff$ value that greatly exceeds current bounds. The asymmetric reheating mechanism \cite{Chacko:2016hvu, Craig:2016lyx} resolves this issue by introducing an additional heavy \emph{reheaton} with sufficiently long lifetime, 
causing it to initiate a period of early matter domination after it decouples from the thermal bath. It then decays out of equilibrium with a larger branching ratio to SM particles, leaving the twin sector at a much lower temperature than the visible sector, and thus suppressing $\dneff$. 
In order to successfully realize asymmetric reheating, the following conditions must be satisfied: 
\begin{itemize}
    \item 
    The temperature of the visible SM bath after it has been reheated by the reheaton decay, $T_{A,R}$, must be below the $\sim$ GeV temperature at which the Higgs portal interaction between the visible and twin sectors decouples, otherwise the generated temperature difference is washed out.
    \item $T_{A,R}$ should be at least an order of magnitude larger than the reheating temperature $T_{B,R}$ in the twin sector to avoid current $\Delta N_{\rm eff}$ bounds (see below for the precise calculation).
    \item 
    The decay of the reheaton must not modify the successful predictions of BBN in the visible sector. 
    If baryogenesis is not related to asymmetric reheating and instead arises due to some higher-scale processes, then the temperature of the SM bath just \emph{before} the asymmetric reheating must never go below $\sim$ MeV, \emph{unless} the subsequent $T_{A,R}$ is so high that it completely erases any memory of the first round of BBN.
    On the other hand, if the baryon asymmetry is generated during asymmetric reheating, then there are no SM baryons to undergo BBN prior to the reheaton decay, and only the SM bath temperature just \emph{after} the asymmetric reheating has to lie above the BBN temperature. 
    Since our baryogenesis model realizes the latter possibility, we only impose the less constraining requirement $T_{A,R} \gtrsim (1 - 10)~\mathrm{MeV}$. 
\end{itemize}
There are different viable candidate scenarios for asymmetric reheating~\cite{Chacko:2016hvu, Craig:2016lyx, Ireland:2022quc}, and we focus our attention on the $\nu$MTH framework~\cite{Chacko:2016hvu}.
The $\nu$MTH model is an extension of the original MTH model that introduces heavy right-handed neutrinos in each sector.
These are singlets under all symmetry groups except for the twin $\mathbb{Z}_2$ symmetry. In a toy one-generation model, the new interactions are  
\begin{gather}
    \mathcal{L} \supset - y_A H_A \bar{L}_A N_A -  y_B H_B \bar{L}_B N_B - \frac{1}{2}M_N(\bar{N}_A N_A^c+\bar{N}_B N_B^c)-M_{AB}\bar{N}_A N^c_B\,,
    \label{eqn:nuMTH}
\end{gather}
where $M_N$ and $M_{AB}$ are Majorana mass terms for the RHNs. 
The $\mathbb{Z}_2$ symmetry forces the Yukawa couplings to obey $y_A = y_B = y$. 
In the simplest scenario, the active neutrino masses are generated via a standard type-1 seesaw mechanism. In the one-generation model, the mass eigenstates and  eigenvalues are given by 
\begin{gather}
    N_+=\frac{1}{\sqrt{2}}\left(N_{A} + N_B\right), \quad N_-=\frac{1}{\sqrt{2}}\left(N_{A} - N_B\right),
    \label{Eqn:transform}\\
m_{N_\pm}=M_N \pm M_{A B}.
\label{e.MNpm}
\end{gather}
Cosmological bounds on active-sterile neutrino mixing require $M_{AB}/M_N \lesssim 10^{-3} - 10^{-2}$~\cite{Hannestad:2012ky}. 
This implies that the $N_+$ and $N_-$ mass eigenstates are almost degenerate and have the same decay widths for all reheating processes. Thus, we are safe to treat them as duplicated mass eigenstates for the purpose of reheating (but not for baryogenesis due to different signs in their couplings, as we will see).

In a realistic three-generation model, 
the above Yukawa term becomes $y_{ki} H  \bar L_k  N_i$. After electroweak symmetry breaking, and denoting with a tilde the left- and right-handed neutrino states most aligned with their respective mass bases before active-sterile mixing, this becomes
$\tilde y_{ki} \frac{v}{\sqrt{2}} \bar {\tilde \nu}_{L_k} \tilde N_i$. 
In general, this $\tilde y_{ij}$ matrix is not diagonal, and each $\tilde N_i$'s lifetime is determined by the sum $\sum_k|\tilde y_{ki}|^2$, generically leading to all right-handed neutrinos contributing to the reheating.
However, we consider the limit in which a single mass eigenstate alone, without loss of generality taken to be the heaviest $\tilde N_1$, serves as the reheaton that dominates the energy density of the universe before it decays out-of-equilibrium. 
This requires $\sum_k|\tilde y_{k1}|^2 \ll \sum_k|\tilde y_{k2}|^2, \sum_k|\tilde y_{k3}|^2$, which is most plausibly achieved if the $\tilde y_{ij}$ matrix is diagonal with elements $y_i$, and $y_1 \ll y_2, y_3$.
The model-building requirements to achieve this separation of masses while respecting the observed active neutrino mixing angles are discussed in detail in~\cref{sec:flavor_alignment}. For now, we  assume that these conditions are satisfied, and implicitly work in the above mass-basis-aligned basis while dropping the tildes from hereon for simplicity.

As we explain in \cref{sec:2-gen-model}, baryogenesis requires $m_{N_1} > 125~\mathrm{GeV}$.
The required lifetime of $N_1$ then implies an extremely small Yukawa coupling, and hence small corresponding active neutrino mass $m_{\nu_1} \approx y_1^2 v^2/m_{N_1} \ll 0.1~\mathrm{eV}$. 
The lighter mass eigenstates $N_{2,3}$ and their respective Yukawas must then generate active neutrino masses $m_{\nu_{2,3}} \sim \mathcal{O}(0.1~\mathrm{eV})$ to reproduce the observed neutrino oscillation pattern~\cite{Denton:2022een, Esteban:2020cvm}, implying a large mass hierarchy between the lightest and the heavier two active neutrino species. 
This also ensures that $N_{2,3}$ play no important role in asymmetric reheating. We now explore this scenario in detail.

Following the discussion in \cite{Chacko:2016hvu}, we first explicitly show that for an unmodified MTH theory, $\dneff$ significantly exceeds the observational bounds. The SM sector and the twin sector interact through the Higgs portal, with an interaction rate and therefore the decoupling temperature that depends on the ratio of the electroweak VEVs $f/v$. For $f/v=3-7$, the decoupling temperature is around $T_D\approx 2-4.4\, \mathrm{GeV}$ \cite{Curtin:2021alk}. After decoupling, the temperatures in the A and B baths evolve separately. Using conservation of comoving entropy, we estimate the twin sector contributions to $\dneff$ as 
\begin{gather}
    \dneff=3  \left.\frac{\rho_B}{\rho_\nu}\right \vert_{\mathrm{BBN}} \approx 7.4   \left.\frac{\rho_B}{\rho_A}\right \vert_{\mathrm{BBN}} = 7.4 \left(\frac{g_{*B, D}}{g_{*A, D}}\right)^{4/3} ,
\end{gather}
where $g_{*A, D} \equiv g_{*A}\vert_{T = T_D}$, similarly for $g_{*B, D}$.
In the case where $f/v=7$, one finds $\dneff\approx 6$, which is much larger than the current experimental bound $\dneff \leq 0.23$ \cite{Planck:2018vyg}.

The asymmetric reheating mechanism fixes this by lowering the hidden sector temperature relative to the SM bath.
As discussed above, we choose the coupling of the heaviest mass eigenstate $N_1$ to be small so that it is long-lived and decays into both A and B sectors with a relative branching ratio
\begin{gather}
    \epsilon\equiv \frac{\Gamma_{{N_1}\rightarrow B}}{\Gamma_{{N_1}\rightarrow A} + \Gamma_{{N_1}\rightarrow B}} \ll 1.
\end{gather}
This ratio has different characteristic size depending on $m_{N_1}$:
\begin{itemize}
\item When the mass of $N_1$ is smaller than the visible $W$ boson mass, $m_{N_1} < m_{W_A}$, its decays in both sectors are all three-body. The decay width into the visible sector is given by
\begin{gather}
\label{eq:decay1}
    \Gamma_A= \frac{y_1^2 m_{N_1}^3}{12288 \pi^3} \left( \frac{g_w^2}{m_{W_A}^2} + \frac{g_z^2}{m_{Z_A}^2} \right) + \frac{3m_{b_A}^2 y_1^2 m_{N_1}^5}{1440 \pi^3(  v^2 m_{H_A}^4) },
\end{gather}
where the last term represents the dominant three-body decay via an off-shell Higgs boson to two bottom quarks and a neutrino, if kinematically accessible. 
The decay width into the twin sector is  obtained by replacing the SM particle masses with their corresponding twin particle masses and the SM VEV $v$ with the twin VEV $f$. 
This decay rate is dominated by the off-shell electroweak gauge boson contributions, since that decay proceeds through an active-sterile neutrino mixing, which introduces extra powers of $v$ and hence reduces the $(m_{N_1}/m_{W,Z})$ suppression.  
As a result, $\epsilon \approx v^2/f^2 \ll 1$ due to the different gauge boson masses in the off-shell propagators, and the dependence on $m_{H_B}$ is negligible. This is the regime that was examined in~\cite{Chacko:2016hvu, Curtin:2021alk}.

\item When $m_{W_A}\simeq m_{Z_A}<m_{N_1}<m_h$, the decay in the twin sector remains three-body while the decay in the visible sector is dominated by the two-body processes involving W and Z bosons
\begin{gather}
\label{eq:decay2}
    \Gamma_A\simeq\frac{y_1^2 m_{W_A}^2}{128 \pi}\left(\frac{m_{N_1}}{m_{W_A}^2}-\frac{3 m_{W_A}^2}{m_{N_1}^3}+\frac{2 m_{W_A}^4}{m_{N_1}^5}\right)+\frac{y_1^2 m_{Z_A}^2}{128 \pi}\left(\frac{m_{N_1}}{m_{Z_A}^2}-\frac{3 m_{Z_A}^2}{m_{N_1}^3}+\frac{2 m_{Z_A}^4}{m_{N_1}^5}\right).
\end{gather}
This greatly enhances the $\epsilon$ suppression.

\item When $m_h<m_{N_1}<m_{W_B}$, decays into the visible Higgs become two-body, adding $\Gamma (N_1 \to h_A \nu_1) = y_1^2 m_{N_1}/16\pi$ to the two-body decay rate in \cref{eq:decay2} and further suppressing the branching ratio into the twin sector. 

\item When $m_{W_B} \ll m_{N_1}$, two-body decays into the twin $W$ bosons open. The decay width into the hidden sector is similar to \cref{eq:decay2}, with all masses replaced by their twin counterparts for available channels. In this regime, the dramatic two-body versus three-body suppression of $N_1$ branching ratios is lost, rendering the asymmetric reheating mechanism ineffective unless $m_{W_B} \sim m_{N_1}$, in which case the 2-body decays in the $B$-sector are still phase-space suppressed.
\end{itemize}

In all cases, the Yukawa coupling of $N_1$ can be chosen to ensure a reheating temperature $T_{A,R}$ below the twin decoupling temperature but above the BBN scale. 
The corresponding energy densities of the two sectors after $N_1$ has decayed can be estimated as
\begin{align}
\rho_{A,R} &= 3 \Gamma_{N_1}^2M_{Pl}^2 = \frac{\pi^2}{30} g_{*A,R} T_{A,R}^4,
\label{eq:rhoA}\\
\rho_{B,R} &= 3\epsilon \Gamma_{N_1}^2M_{Pl}^2 + \left(\frac{g_{*B,D}}{g_{*B,R}}\right)^{1/3} \rho_{B,D} \left(\frac{a_D}{a_R}\right),
\label{eq:rhoB}
\end{align}
where the $D$ subscript on the effective numbers of relativistic species and scale factors $a$ indicates they are evaluated when $T_A = T_B = T_D$, while $R$ subscripts indicate evaluation when $T_A = T_{A,R}$ immediately after asymmetric reheating via $N_1$ decay. The right-hand side of \cref{eq:rhoA} comes entirely from the $N_1$ relic density, since it dominates the energy density of the universe prior to decaying. This gives 
\begin{equation}
    T_{A,R}  \ \propto  \  y_1 \ g_{*A,R}^{-1/4} \ .
\end{equation}
In the B sector, the energy dump from $N_1$ decays in the first term of \cref{eq:rhoB} does not necessarily dominate the energy density after reheating due to suppression from the small branching ratio $\epsilon$. Therefore, we cannot neglect the primordial energy density of the B sector, given by the second term of \cref{eq:rhoB}.The ratio between \cref{eq:rhoB} and \cref{eq:rhoA} can be rewritten as
\begin{equation}
    \frac{\rho_{B,R}}{\rho_{A,R}} = \epsilon +R_N \, ,
\end{equation}
which defines $R_N$ as the ratio of the primordial energy density in the B sector to the total energy dump of $N_1$ decays into both sectors. With this, $\dneff$ is given by
\begin{gather}
    \label{e.epsilonRN}
    \dneff=7.4\left(\frac{g_{*B}}{g_{*A}}\right)^{1/3}(\epsilon+R_N).
\end{gather}
This showcases that in order to effectively realize asymmetric reheating, both $\epsilon$ and $R_N$ must be small.

\subsection{$\dneff$ and twin BBN in the $\nu$MTH with single dominant reheaton}

Let us now delineate how $\dneff$ depends on the parameters of the $\nu$MTH model in our scenario where $N_1$ dominates reheating, and compare the predictions to current cosmological bounds. 
As we discuss below, generating a CP asymmetry for baryogenesis requires an on-shell Higgs to be present in the reheaton decay chain, which means that $m_N > 125~\mathrm{GeV}$. This leads us to examine a region of parameter space different to the one considered in existing studies, which focused on GeV-scale right-handed neutrinos. 

\begin{figure}
    \centering
    \includegraphics[scale=0.55]{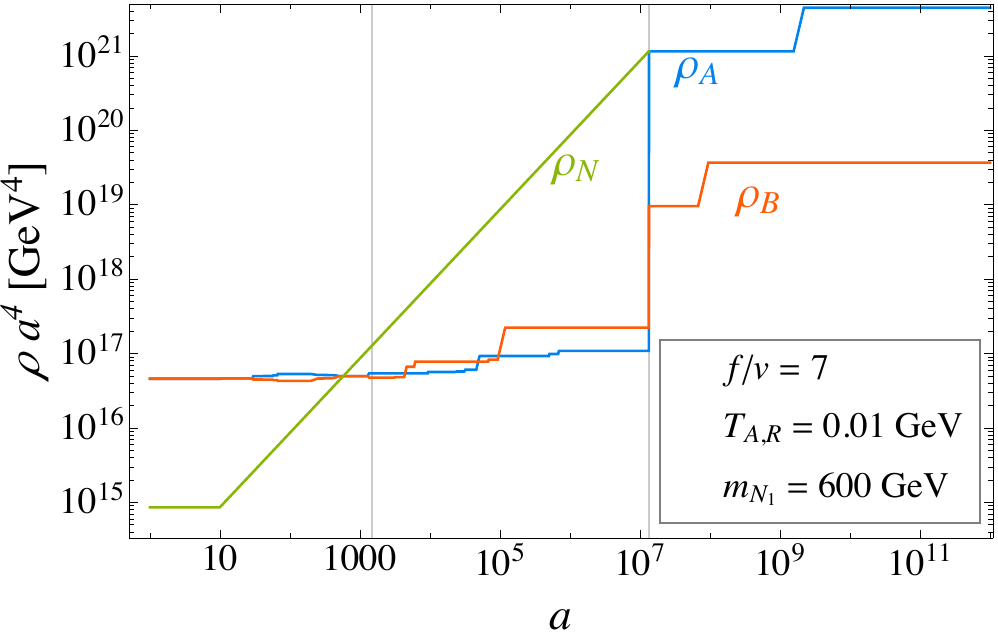}
    \caption{Evolution of energy densities of the photon bath ($\rho_A$), twin photon bath ($\rho_B$) and reheaton ($\rho_N$) for $f/v=7$, $m_N=600$~GeV and a reheating temperature of $10$~MeV. The horizontal axis shows the scale factor $a$, and the vertical axis shows $\rho a^4$. The left vertical line marks the thermal decoupling of the visible and twin sectors while the right vertical line marks reheating. Note that $(\rho_{A} + \rho_{B})a^4$ is constant prior to Higgs decoupling, with transient departures from equipartition only occurring when $g_*$ differs between the two sectors due to different mass thresholds.
    }
    \label{fig:rhoplot}
\end{figure}

To illustrate the cosmological history of the mechanism, \cref{fig:rhoplot} shows the evolution of the energy densities in the SM and twin sectors, as well as that of $N_1$. The scale factor is normalized to unity at $T=10\,m_{N_1}$ for convenience. 
We work in the simplifying instant decay/annihilation approximations, which should not significantly affect our results.
Conservation of entropy therefore gives a relation between the temperatures before and after a mass threshold,
\begin{gather}
    g_{\star S}(T_i) T_i^3= g_{\star S}(T_j) T_j^3 \ ,
\end{gather}
where $g_{\star S}$ is the effective number of degrees of freedom in entropy. 
Although it gives a very good approximation for the subsequent temperature and density evolution, an instantaneous temperature increase at the threshold is unphysical.
We therefore replace the artificial temperature discontinuities by flat temperature plateaus, which are more representative of the actual period of slow logarithmic cooling during entropy injections.
Following the green line in \cref{fig:rhoplot}, the $N_1$ become non-relativistic around $a \sim 10$ and dominate the energy density of the universe starting at $a\sim 10^3$ until they decay around $a\sim 10^7$, raising the temperature and thus the energy density of the SM (blue line) preferentially compared to the twin sector's energy density (orange line).

As discussed in the previous subsection,  $\dneff$ only depends on the SM bath temperature $T_{A,R}$ immediately after $N$ decays
, the mass of $N_1$, and $f/v$. In our regime of interest with strong asymmetric reheating, $m_{N_1} < m_{W_B}$,  the $H_B$ mass has a negligible effect.  For concreteness, we set $m_{H_B} = 1.5~\mathrm{TeV}$ in our plots, but the results apply for any TeV-scale  twin Higgs mass.

\begin{figure}
    \centering
    \hspace*{-9mm}
\begin{tabular}{lll}
\includegraphics[width=.35\linewidth]{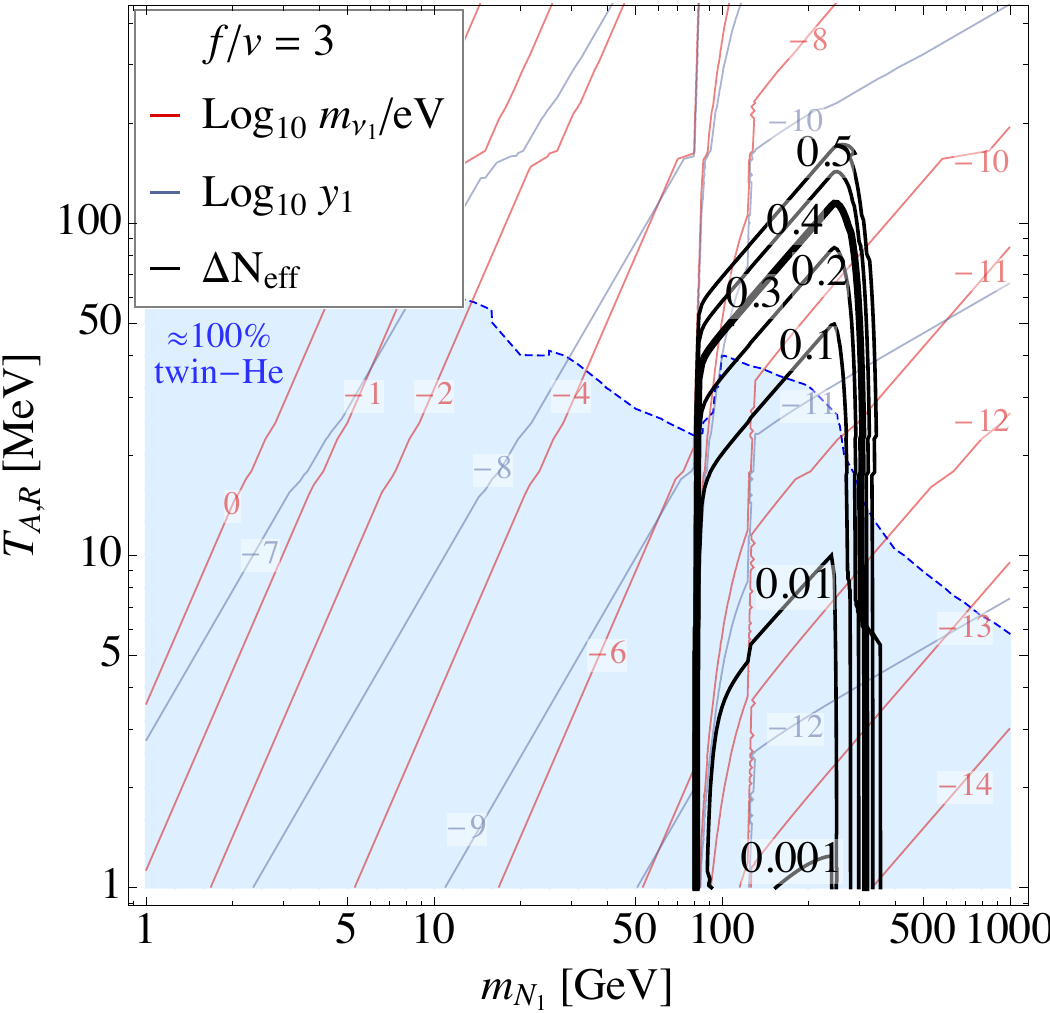} & 
\hspace*{-4mm}
\includegraphics[width=.35\linewidth]{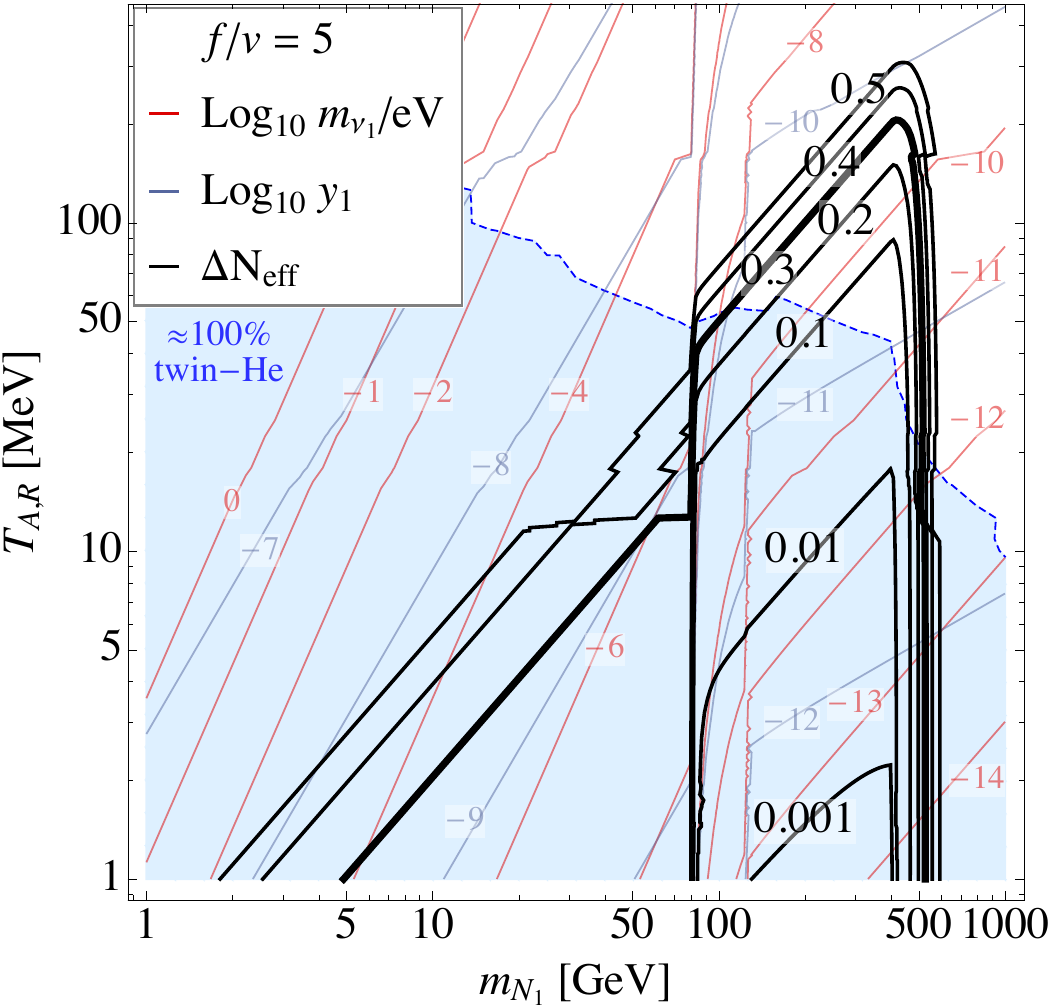} & 
\hspace*{-4mm}\includegraphics[width=.35\linewidth]{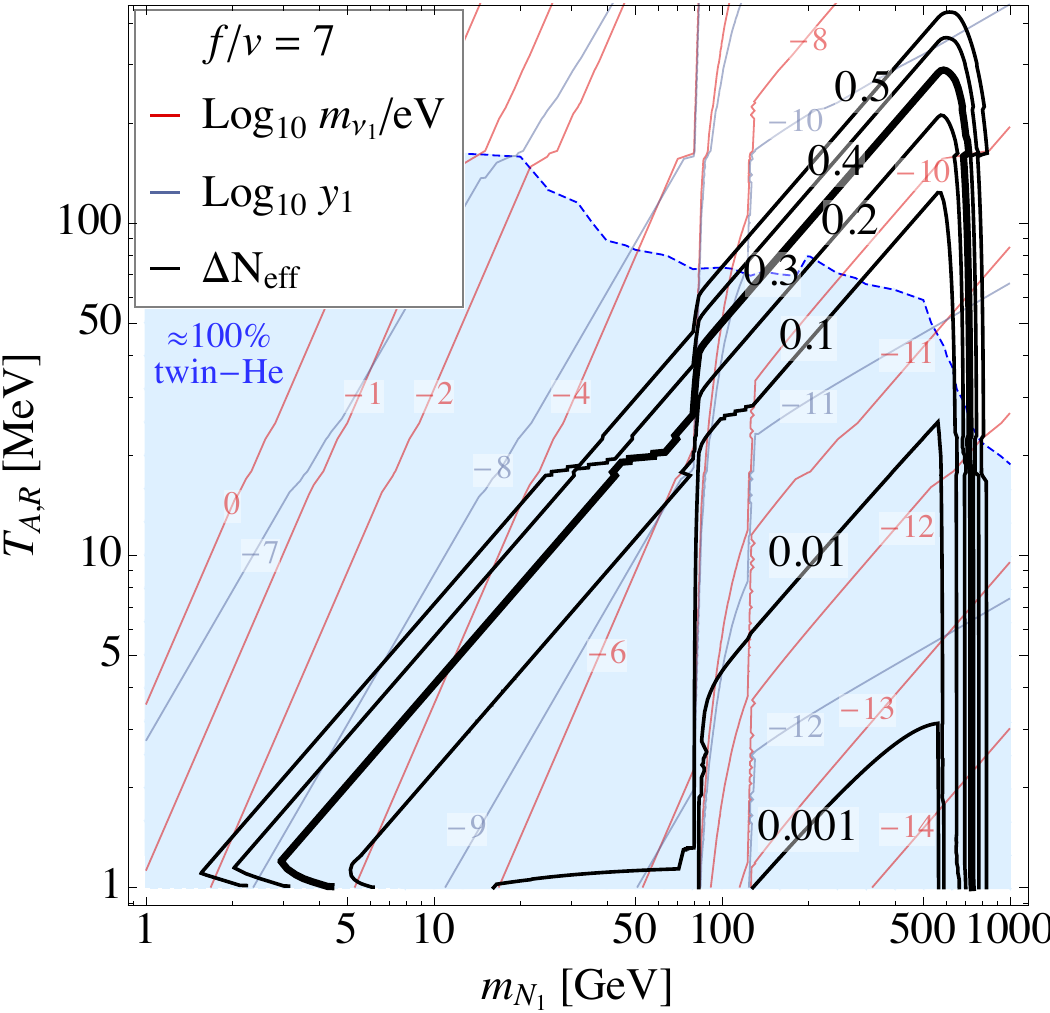}
\end{tabular}
\caption{
Black contours show predictions for $\dneff$
as a function of the $N_1$ mass and reheating temperature $T_{A,R}$ in the visible sector, for the $\nu$MTH model with $f/v=3,5,7$ and a single dominant reheaton $N_1$. The $\dneff = 0.3$ contour is thicker to highlight the region consistent with the Planck bound on dark radiation~\cite{Planck:2018vyg}.
Blue contours show the Yukawa coupling $y_1$ set by $T_{A,R}$, with red contours showing the corresponding lightest active neutrino mass.
The blue shaded region indicates where $T_{B,R}$ is so low that twin-neutron-proton weak transitions do not equilibrate, such that the roughly equal amounts of twin neutrons and protons produced in $N_1$ decays combine to $\approx 100\%$ twin Helium. Twin BBN proceeds normally~\cite{Bansal:2021dfh} above the blue shaded region. 
}
    \label{fig:dneff}
\end{figure}

Contours of $\dneff$ for different values of $f/v$ in the $T_{A,R}$ vs.~$m_{N_1}$ plane are shown in \cref{fig:dneff}. The regions of parameter space consistent with the latest Planck bounds on dark radiation~\cite{Planck:2018vyg} are highlighted with the thicker $\dneff = 0.3$ contour.
Although we display a wide mass range for illustration purposes, we are particularly interested in the $m_h < m_{N_1} \lesssim m_{W_B}$ subrange.
The maximal value of $m_{N_1}$ that can accommodate the $\dneff$ bounds strongly depends on $f/v$.
This upper bound on the reheaton mass has important implications for our baryogenesis mechanism, which requires the addition of new colored states in the decay of $N_1$.

\subsection{Twin baryons and atomic dark matter}

The properties of the twin baryons generated in our setup follow from the lack of hard $\mathbb{Z}_2$ breaking, which means that the only difference between the two sectors is the different Higgs vevs. This results in different fundamental fermion masses by a factor of $f/v$, as well as modifications in the RG running, most importantly for the strong coupling constant in the twin sector.
Twin baryons therefore realize a particular atomic dark matter scenario, with SM-QED-like dark coupling $\alpha_D = \alpha_{em}$, dark electron mass $m_{e_D} = (f/v) m_{e}$, and dark proton mass $m_{p_D} \sim (1.3 - 1.7) m_{p}$, as well as dark nuclear interactions. The dark proton mass is derived from the ratio of twin to visible confinement scales~\cite{Chacko:2018vss}, which is approximately\footnote{
While we use \cref{e.mpBovermpA} for consistency with the cosmological analysis of~\cite{Bansal:2021dfh}, this estimate  has been superseded by the more sophisticated lattice-based estimates of~\cite{Hippert:2021fch}. However, the two estimates are within 10\%, so this will not significantly affect our results.}
\begin{equation}   
    \label{e.mpBovermpA}
    \frac{m_{p_B}}{m_{p_A}}\approx 0.68+0.41\log\left(1.32+\frac{f}{v}\right) \, .
\end{equation}

Twin BBN for the $\mathbb{Z}_2$-symmetric MTH was most recently studied in~\cite{Bansal:2021dfh}, finding large twin helium mass fractions of $\sim 0.75$ for $\dneff \sim 0.1$. This is due to a larger neutron-proton freeze-out ratio in the twin sector compared to the SM.
Since the twin neutron is significantly longer-lived than the SM neutron, this $(n_B/p_B)$ ratio survives almost unchanged until $T_B$ drops below the twin-deuterium bottleneck $T_{B,D}$ and is almost entirely converted to twin Helium.

However, twin BBN is modified if the reheating temperature in the twin sector is so low that twin-neutron-proton weak transitions do not equilibrate. This is the case in the blue shaded regions of \cref{fig:dneff}. The twin Helium abundance is therefore entirely given by the $(n_B/p_B)$ ratio generated by the hadronization of the reheaton decay products, rather than the freeze-out ratio.\footnote{In this whole region, $T_{B,R}$ is still either above or only slightly below $T_{B,D}$, meaning twin deuterium fusion is still efficient after reheating of the twin sector.} 
In all the models that we study, which feature baryogenesis from the decay of reheatons with mass $m_{N_1} \gg \mathrm{GeV}$, 
isospin is an excellent symmetry of hadronization, implying equal production of neutrons and protons~\cite{ParticleDataGroup:2022pth}.
As a result, almost all the twin baryons will be converted to twin helium. The blue shaded region therefore corresponds to pure twin helium atomic dark matter, realizing an interesting scenario that was analyzed e.g. in~\cite{Chacko:2021vin} for direct detection. It is also noteworthy that very small $\dneff \sim 0.01$ always results in pure-twin-helium atomic dark matter, and this conclusion should model-independently apply to other asymmetric reheating mechanisms as well. 

The twin baryon fraction strengthens cosmological bounds on $\dneff$ due to the presence of dark baryo-acoustic oscillations. The recent study~\cite{Bansal:2021dfh} used Planck and BAO data to obtain updated bounds on $\dneff$ within the MTH for different $f/v$ and twin baryon dark matter fraction $\hat{r} = \Omega_{aDM}/\Omega_{DM}$.
For $\hat r \lesssim 0.03$, the standard  Planck bound~\cite{Planck:2018vyg} of $\dneff \sim 0.25-0.3$ is valid for $f/v=3-7$, as the twin BAOs have minimal additional effect beyond that of the dark radiation. For larger twin baryon fractions, twin BAOs start to more severely constrain the temperature in the twin sector, translating to tighter effective bounds on $\dneff \lesssim 0.1 - 0.2$. 
We take these more severe bounds into account in our subsequent analysis of twin baryogenesis.\footnote{The bounds of \cite{Bansal:2021dfh} were obtained under the assumption that twin BBN proceeds normally, which is not true in the blue shaded regions of~\cref{fig:dneff}. This does not affect our results, as twin baryogenesis in our model is only effective for reheating temperatures high enough for standard twin BBN.}

In the later universe, these twin baryons can have a large variety of astrophysical signatures. 
As a realization of atomic dark matter, their dissipation and collapse can impact galactic dynamics~\cite{Fan:2013yva,Fan:2013tia,McCullough:2013jma,Randall:2014kta,Schutz:2017tfp, Buch:2018qdr, Ghalsasi:2017jna, Ryan:2021dis, Gurian:2021qhk, Ryan:2021tgw, Foot:2013lxa,  Foot:2014uba, Foot:2015mqa,Chashchina:2016wle, Foot:2017dgx, Foot:2018dhy, Foot:2016wvj, Foot:2013vna}, with recent N-body simulations exploring these effects in the milky way and dwarf satellites from first principles~\cite{Roy:2023zar, Gemmell:2023trd}.\footnote{Directly applying these studies of minimal aDM to the Twin Higgs is not straightforward, since twin baryons give rise to feedback processes that are absent in minimal aDM.}
They collapse to form mirror stars~\cite{Curtin:2019ngc, Curtin:2019lhm, Howe:2021neq, Winch:2020cju}, mirror white dwarfs~\cite{Ryan:2022hku} mirror neutron stars~\cite{Hippert:2021fch, Hippert:2022snq}
and black holes with non-standard masses~\cite{Pollack:2014rja, Shandera:2018xkn, Singh:2020wiq, Gurian:2022nbx, Fernandez:2022zmc}, which could be detected and distinguished from SM astrophysical objects in gravitational wave observations, microlensing surveys, and electromagnetic emissions if mirror stars capture SM matter via photon kinetic mixing interactions. 
Our concrete theoretical realization of twin baryogenesis in the $\mathbb{Z}_2$ symmetric MTH model thus further motivates searching for these spectacular astrophysical signals.


\section{The $\nu\phi$MTH Model}
\label{sec:2-gen-model}
We now extend the $\nu$MTH for visible and twin baryogenesis during decay of the dominant reheaton. The main new ingredient is a new colored scalar $\phi$ in both SM and twin sectors that allows the RHNs to couple to the quark sector. Two main cases for the coupling hierarchies are presented: a generation universal and a top-philic scenario. For each case, we discuss the current bounds on the strengths of these couplings from collider searches and proton decay experiments, and highlight the required flavor alignments in the colored scalar and the right-handed neutrino couplings. We end the section by discussing why, in the $\nu$MTH, the RHN mass eigenstate that does the asymmetric reheating has to be the same state that generates the baryon asymmetries, setting the stage for the baryogenesis calculations of the next sections.

\subsection{Model Definition}
\label{sec:model_defn}

The minimal extension of the $\nu$MTH model that allows a small fraction of $N$ decays to generate a baryon asymmetry involves adding a color-triplet scalar to both sectors in a $\mathbb{Z}_2$-invariant way. These scalars $\phi_{A,B}$ have (twin) hypercharge $Y_\phi = +2/3$ or $Y_\phi= -1/3$, meaning they have the quantum numbers of a right-handed up- or down-type squark. Since no hard $\mathbb{Z}_2$ breaking is assumed, the couplings of $\phi_{A,B}$ are the same in both sectors. In the $Y_\phi = +2/3$ case, the non-kinetic terms in the Lagrangian are given by 
\begin{equation}
    \mathcal{L} = \mathcal{L}_{\nu\textrm{MTH}} + \mathcal{L}_{\phi,\text{int}}- V,
\end{equation}
where
\begin{equation}
    \mathcal{L}_{\phi,\text{int}}^{Y_\phi = 2/3} = - 
    \lambda_{mi} \phi^{}_A  u_{m,A}^c \bar{N}_{i,A} - \lambda'_{mn} \phi_A^\dagger \bar{d}^{}_{m,A} d_{n,A}^c
    +
    (A \to B)
    \label{eq:L_phi_int}
\end{equation}
and
\begin{equation}
    \label{e.fullpotential}
    \begin{aligned}
   V &= \mu_\phi^2 (|\phi_A|^2+|\phi_B|^2) + \mu^2 (|H_A|^2+|H_B|^2)\\ 
   &+\sigma(|H_A|^2+|H_B|^2)^2 + \sigma_\phi (|\phi_A|^2+|\phi_B|^2)^2\\
   &+\xi(|\phi_A|^2\left|H_A\right|^2 + |\phi_B|^2\left|H_B\right|^2) + \xi' (|\phi_A|^2\left|H_B\right|^2 + |\phi_B|^2\left|H_A\right|^2).
    \end{aligned}
\end{equation}
Here, $i,j$ denote RHN generation indices and $m,n$ are quark generation indices.  Note that the $\lambda'$ term has a color structure proportional to the anti-symmeteric tensor $\epsilon_{abc}$, and thus flavor indices on the down type quarks must be antisymmetrized as well: down-type quarks in the second and fourth term in \cref{eq:L_phi_int} must be from different generations. 
$\mathcal{L}_{\nu \mathrm{MTH}}$ denotes the 3-generation generalization of \cref{eqn:nuMTH}. 
Going to the $\mathbb{Z}_2$-eigenstate basis defined in  \cref{Eqn:transform}, the terms involving the right handed neutrinos become:
\begin{alignat}{2}
\label{e.LY23}
\begin{aligned}
    & \mathcal{L}_{\nu \mathrm{MTH}}  & =  & -\frac{y_i}{\sqrt{2}} \left[(\bar{L}_{A,i} H_A+\bar{L}_{B,i} H_B)N_{+,i}  +(\bar{L}_{A,i}H_A-\bar{L}_{B,i}H_B)N_{-,i}\right]  - M_{N_i} (N_{+,i}^2  + N_{-,i}^2), \\  
    & \mathcal{L}_{\phi,\text{int}}^{Y_\phi = 2/3} & =   & -\frac{\lambda_{mi}}{\sqrt{2}} \left[(\phi_A u_{m,A}^c +\phi_B u_{m,B}^c) \overline{N}_{+,i}+(\phi_A u_{m,A}^c -\phi_B u_{m,B}^c) \overline{N}_{-,i}\right]\\
    & & & -\lambda'_{mn}  \left[\phi_A^\dagger \bar{d}^{}_{m,A} d_{n,A}^c +  \phi_A^\dagger \bar{d}^{}_{m,B} d_{n,B}^c\right],
\end{aligned}
\end{alignat}
where $N_{\pm,i}$ are $i$-th generation $\mathbb{Z}_2$ even/odd mass eigenstates respectively, and we have explicitly assumed that the mass difference between the even and odd states is negligible. We also assume diagonal and hierarchical $y_i$, as explained below \cref{e.MNpm} and in \cref{sec:flavor_alignment}.

For the case of $Y_\phi = -1/3$, only the scalar interaction part of the Lagrangian changes. In the flavor basis, we have
\begin{gather}\label{eq:L_phi_13}
\begin{aligned}
     \mathcal{L}_{\phi,\text{int}}^{Y_\phi = -1/3}= -  \lambda_{mi} \phi^{}_A  d_{m,A}^c \bar{N}_{i,A} - \lambda'_{mn} \phi_A^\dagger \bar{u}^{}_{m,A} d_{n,A}^c +  \lambda''_{mn} \phi_A^\dagger \bar{Q}^{}_{m,A} Q_{n,A}^c  + (A \to B).
\end{aligned}
\end{gather}
Moving to the $\mathbb{Z}_2$-eigenstate basis before active neutrino mixing, \cref{eq:L_phi_13} becomes
\begin{alignat}{2}
\label{e.LY13}
\begin{aligned}
    \mathcal{L}_{\phi,\text{int}}^{Y_\phi = -1/3} \supset &-\frac{\lambda_{mi}}{\sqrt{2}}  \left[(\phi_A d _{m,A}^c + \phi_B d_{m,B}^c) \overline{N}_{+,i}+(\phi_A d_{m,A}^c -\phi_B d_{m,B}^c) \overline{N}_{-,i}\right]\\
    &- \lambda'_{mn} \phi_A^\dagger \bar{u}^{}_{m,A} d_{n,A}^c - \lambda'_{mn} \phi_B^\dagger \bar{u}^{}_{m,B} d_{n,B}^c \\
    & +  \lambda''_{mn}  
    \phi_A^\dagger\bar{Q}^{}_{m,A} Q_{n,A}^c +  \lambda''_{mn} \phi_B^\dagger \bar{Q}^{}_{m,B} Q_{n,B}^c .
\end{aligned}
\end{alignat}

In the absence of any explicit $\mathbb{Z}_2$ breaking, any difference in the $\phi_{A,B}$ masses is generated via EWSB through their respective Higgs couplings. When $H_A$ and $H_B$ obtain VEVs, the color triplet scalars acquire masses
\begin{gather*}
    m^2_{\phi_A} = \mu_\phi^2 + \xi v^2 +\xi' f^2,\\
    m^2_{\phi_B} = \mu_\phi^2 + \xi f^2 + \xi'v^2.
\end{gather*}
Defining $\Delta\xi = \xi-\xi'$, the mass of $\phi_B$ can be written in terms of $m_{\phi_A}$ as
\begin{gather}
    \label{e.mphiBdeltaxi}
    m_{\phi_B} = \sqrt{m_{\phi_A}^2 + \Delta\xi \, v^2 (f^2/v^2 -1)}.
\end{gather}
This parametrization will be useful to explore the possibilities of $\phi_B$ being heavier ($\Delta\xi >0$), lighter ($\Delta \xi <0$) or the same mass ($\Delta \xi = 0 $) as its SM counterpart.
Note that in the absence of hard $\mathbb{Z}_2$ breaking, the twin protection for the light Higgs mass is maintained when we add these new degrees of freedom, another important motivation for maintaining the $\mathbb{Z}_2$ symmetry of the MTH. The mass of the colored scalars, while ideally above the TeV scale to satisfy LHC constraints, is therefore not directly tied to the naturalness of the model.

There is potential for rich flavor phenomenology in the couplings of the color triplet scalar to quarks.
This is a topic that has been studied in detail in previous studies dealing with similar particles~\cite{Davidson:1993qk,Giudice:2011ak,Agrawal:2014aoa,Fajfer:2020tqf,Alonso-Alvarez:2021qfd,Alonso-Alvarez:2021oaj}.
We consider two simplified scenarios for the flavor structure of the new $\phi$ couplings to SM quarks:
\begin{itemize}
    \item \textbf{Generation universal couplings}: All quarks couple to the RHN and colored scalar with the same strengths, in both sectors. This implies $\lambda_{mi} = \lambda$ and $\lambda'_{mn} \equiv \lambda'$. For $Y_\phi=-1/3$ this scenario also also implies $\lambda''_{mn} \equiv \lambda''$, but this coupling has no impact on our results in the viable region of parameter space. 
    
    \item \textbf{\emph{Top-philic} couplings}: The RHNs and the colored scalar couple dominantly to the
    the third-generation in both sectors. If the above Lagrangians are taken to be written in the quark mass basis:
\begin{itemize}
    \item For $Y_\phi = 2/3$, this implies that $\lambda_{3i}\neq 0$ and all other $\lambda \approx 0$. For $\lambda'$ we do not require any particular flavor alignment. For definiteness, we choose $\lambda'_{12}\neq 0$ while the other $\lambda'_{mn} = 0$, but other choices do not affect our conclusions.
    \item For $Y_\phi=-1/3$, this means that $\lambda_{3i}, \lambda'_{33}, \lambda''_{33} \neq 0$ and all other $\lambda, \lambda', \lambda''$ couplings are set to zero. 
    Proton decay bounds will generically require $\lambda''$ to be small, which will be discussed in \cref{sec:pd}.
\end{itemize}
\end{itemize}
In the top-philic $Y_\phi = -1/3$ scenario, the required $\lambda''$ suppression may naturally originate from a supersymmetric UV completion of the Twin Higgs.
Such models usually consists of the Minimal Supersymmetric Model (MSSM) and a twin copy MSSM', with an additional singlet superfield to connect the SM and twin Higgs superfields \cite{Falkowski:2006qq,Chang:2006ra, Craig:2013fga,  Katz:2016wtw, Badziak:2017syq, Badziak:2017kjk, Badziak:2017wxn, Asadi:2018abu, Choi:2023eus}. For $Y_\phi=-1/3$, our color triplet interaction $\lambda'_{ij} \phi_A^\dagger \bar{U}^{}_{i,A} D_{j,A}^c$ can be identified with an R-parity violating (RPV)~\cite{Barbier:2004ez} supersymmetric term, where we recognize $\phi_{A,B}$ as visible- and twin-sector bottom squarks. The $\lambda''$ terms are however forbidden in the superpotential due to holomorphy. They are only generated at one loop through an RPV term and the quark Yukawa terms from the superpotential. Thus, $\lambda''$ is suppressed by a loop factor and the small non-top Yukawa couplings.
Identifying the new colored scalars in the $\nu$MTH with bottom squarks further implies that $\xi = y_b^2$ in the $\xi|\phi_A|^2\left|H_A\right|^2$ term and its twin counterpart.
In that case, one would expect $\Delta\xi \sim y_b^2 \sim 0$ for our purposes, and $\mu_\phi^2$ to be around the scale of  soft supersymmetry breaking.
However, the details of the supersymmetric UV completion can be changed, including the exact identity of the $\phi_{A,B}$ scalars, while maintaining a general expectation of small $\lambda''$.

The generation-universal scenario and the top-philic scenario yield drastically different phenomenology in their predictions for the generation of a matter-antimatter asymmetry in the visible and twin sectors.
They are also constrained very differently by proton decay bounds, which exclude most of the parameter space of the generation-universal coupling scenario. For top-philic couplings, proton decay constraints can be avoided, but this relies on quark rotations from the flavor-basis to the mass basis not ruining the top-philic alignment of the $\lambda, \lambda'$ couplings, and $\lambda''$ being small enough to have negligible effect. We discuss this in more detail below.

\subsection{LHC searches for color-triplet scalars}
\label{sec:lhcconstraints}

Being strongly interacting and with a mass that might be near the TeV scale, $\phi_{A}$ has important implications for terrestrial collider searches.
Its possible decay channels are into a diquark final state through  $\lambda^{\prime(\prime)} \bar q  q^c \phi_A^\dagger$ operators, or to a quark, a Higgs boson and a neutrino via an off-shell $N_i$.
The former channel is the dominant one in the region of parameter space of interest to us.

The color-triplet scalar can be pair produced at high-energy colliders like the LHC via the strong interactions, and its diquark decay leads to a four-jet signatures.
Searches for this kind of final state have been performed at ATLAS~\cite{ATLAS:2017jnp} and CMS~\cite{CMS:2018mts,CMS:2022usq} as part of their supersymmetry search programs.
The most recent limits from the CMS analysis of $138\,\mathrm{fb}^{-1}$ of data yields a $95~\%$ CL lower limit of $0.77$~TeV on the mass of $\phi_A$.
It is worth noting however that a narrow mass range $0.52-0.58$~TeV escapes the exclusion and that a $2.5\,\sigma$ (global significance) excess at $0.95$~TeV is found in the data.

The colored scalar $\phi_A$ can also be resonantly produced at the LHC via its diquark $\lambda^{\prime(\prime)}_{mn}$ coupling, leading to a dijet signature.
Searches for this kind of signal at ATLAS~\cite{ATLAS:2019fgd} and CMS~\cite{CMS:2018mgb,CMS:2019gwf} can thus be used to place to limits on the $\lambda'_A$ couplings as a function of the $\phi_A$ mass.
The aforementioned searches present their results for a Gaussian resonance as a function of its mass, and for different relative widths $\Gamma_\phi / m_\phi$.
In order to recast them to our model, we calculate the leading-order production cross-section of $\phi_A$ as a function of $\lambda'_{mn}$ and $m_{\phi_A}$ using \texttt{MadGraph5\_aMC@NLO}~\cite{Alwall:2014hca} with a custom model implemented in \texttt{FeynRules}~\cite{Alloul:2013bka}.
Turning on a single $\lambda'_{mn}$ at a time, the relative width of $\phi$ is given by
\begin{equation}
    \frac{\Gamma_{\phi_A}}{m_{\phi_A}} = 2 \frac{{\lambda'_{mn}}^2}{16\pi} \simeq 1\% \,\left(\frac{{\lambda'_{mn}}}{1}\right)^2.
\end{equation}
For the relevant range of couplings, this intrinsic width is below the experimental resolution of a few percent of the signal width.
We can thus use the narrow width approximation to translate the cross section limits into limits in the $(m_\phi,\,\lambda'_{ij})$ parameter space.
For this, we assume a signal acceptance of $A\sim 0.6$ as appropriate for our isotropic decays, and a branching ratio of $1$ to each $q_i \, q_j$ final state under consideration.

\begin{figure}
\centering
\centerline{
\includegraphics[width=0.5\linewidth]{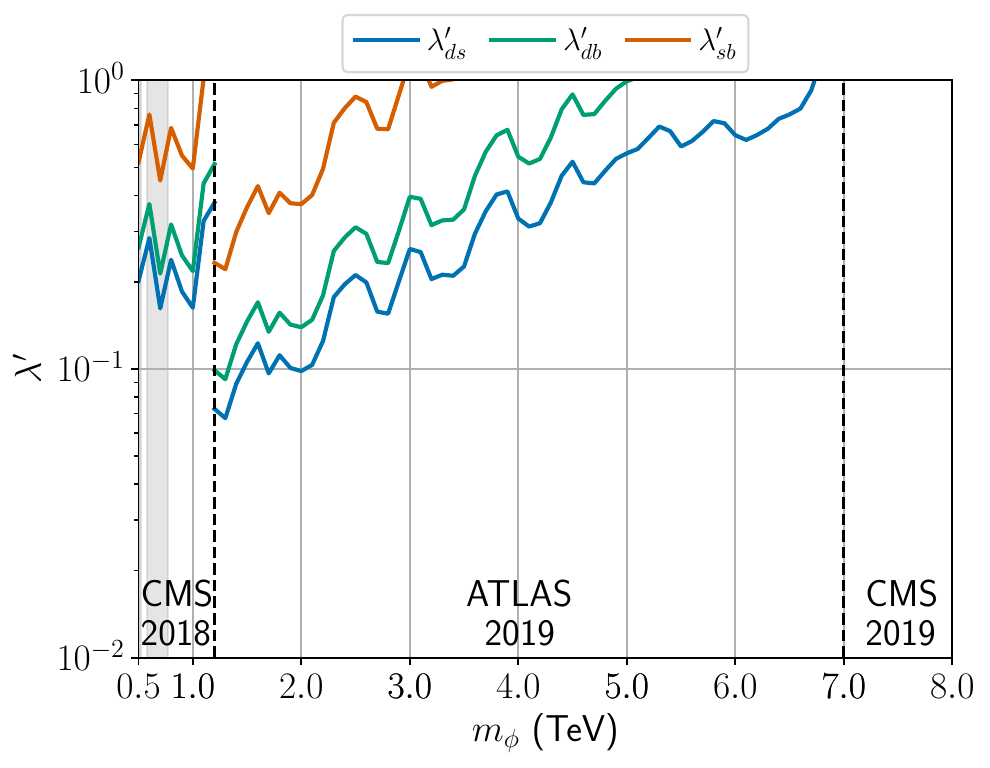}
\includegraphics[width=0.5\linewidth]{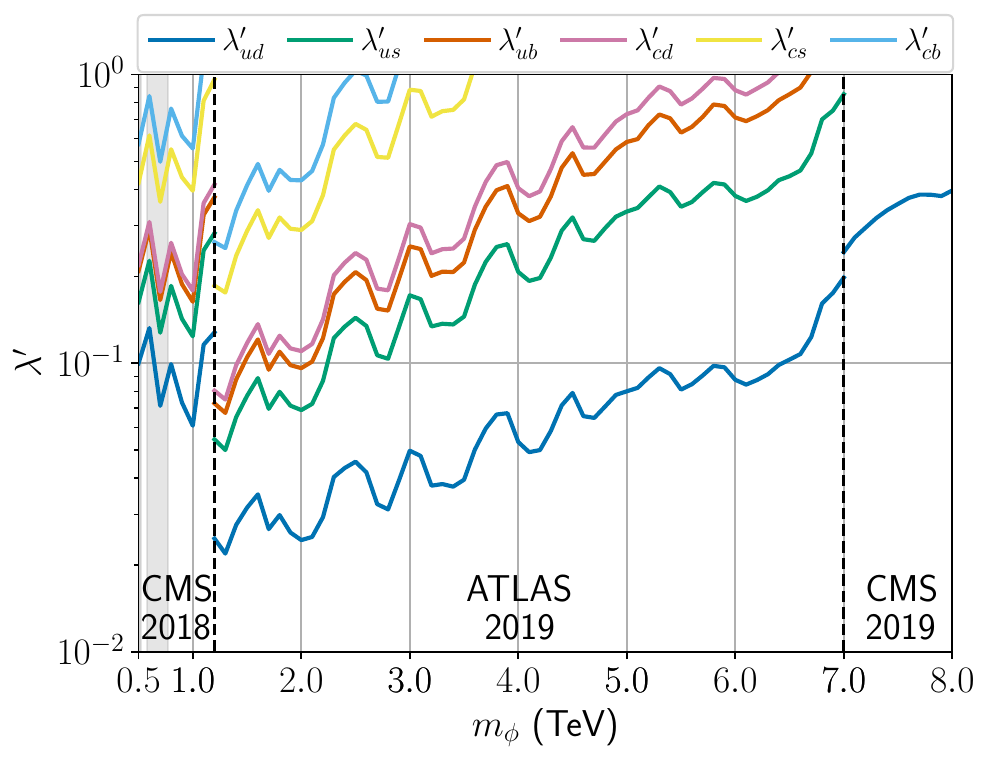}
}
\caption{Collider limits on the diquark couplings of the color-triplet scalar as a function of its mass for resonant single-scalar production (left: hypercharge $2/3$ model, right: hypercharge $-1/3$ model). Four-jet searches~\cite{ATLAS:2017jnp,CMS:2018mts,CMS:2022usq} for pair-produced scalars exclude the mass regions corresponding to the vertical grey bands independently of the diquark coupling.
A collection of dijet searches~\cite{CMS:2018mgb,CMS:2019gwf,ATLAS:2019fgd} for resonantly produced colored scalars exclude the range of couplings above the line for the corresponding flavor variation of the diquark coupling.}
\label{fig:dijet}
\end{figure}

The ensuing limits are shown in \cref{fig:dijet}, left for $Y_\phi = 2/3$ and 
right for $Y_\phi = -1/3$.
In the latter we only show the limits on the $\lambda'$ couplings for clarity: the corresponding $\lambda''$ bounds are quantitatively similar.
The different lines correspond to different flavor variations of the diquark coupling, which are assumed to be turned on one at a time.
As expected, the limits are strongest for light quark couplings, and degrade for heavy flavor ones.
Of relevance for the $Y_{\phi} = -1/3$ top-philic model, we note that the negligible top quark PDF of the proton means that there are no meaningful LHC limits on $\lambda^{\prime(\prime)}_{33}$ from resonant production.

\subsection{Proton decay} \label{sec:pd}

\begin{figure}
    \centering
    \includegraphics[width=0.47\linewidth]{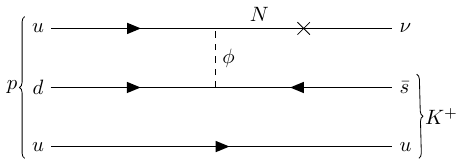}
    \includegraphics[width=0.47\linewidth]{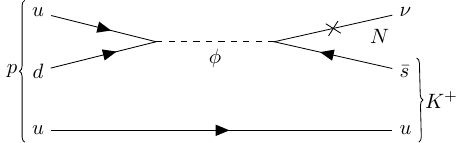}
    \includegraphics[width=0.47\linewidth]{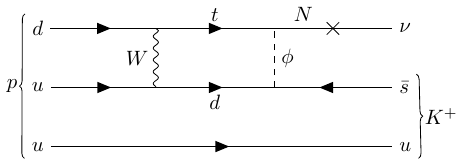}
    \includegraphics[width=0.47\linewidth]{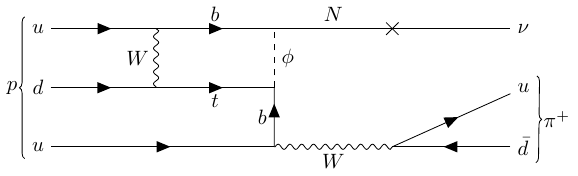}
    \caption{Represeantive diagrams leading to proton decay $p\rightarrow K^+ \nu$ or $p\rightarrow \pi^+ \nu$.
    The left-hand side diagrams correspond to the hypercharge $2/3$ mediator, while those in the right-hand side are for the $-1/3$ one.
    For flavor-universal couplings, the upper two tree-level diagrams are possible, while a top-philic mediator only induces proton decay at one loop via the lower two diagrams.}
    \label{fig:pdecay}
\end{figure}

Since baryon number must be broken in order to allow for the generation of a baryon asymmetry, proton decay is generally predicted in our model.
Proton decay can be induced via the diagrams in \cref{fig:pdecay}.
Since a mass mixing insertion is needed to obtain the final state neutrino, the process is always suppressed by the small neutrino mass.
Thus, we can neglect the contribution to the rates of $N_1$, as its corresponding neutrino $\nu_1$ is in our setup much lighter than the other two neutrino flavors.

Following the prescription in Ref.~\cite{Aoki:2017puj}, the proton decay rate into a meson $M$ and a lepton whose mass is neglected can be written as
\begin{equation}
    \Gamma(p\rightarrow M l) = \frac{m_p}{32\pi} \left[ 1- \left( \frac{m_M}{m_p} \right)^2 \right]^2 \left| C^I W_0^I(M\rightarrow p) \right|^2,
\end{equation}
where $C^I$ is the Wilson coefficient of the effective four-fermion operator inducing the decay, and $W^I_0$ is the corresponding hadronic matrix element.

\paragraph{Generation-universal couplings.}

For generation-universal $\phi_A$ couplings, the leading processes is $p \rightarrow K^+ \nu$ as shown in the top two diagrams of \cref{fig:pdecay}, for the hypercharge $2/3$ and $1/3$ mediators respectively ($p \rightarrow \pi^+ \nu$ has a similar rate but is less constrained experimentally~\cite{Workman:2022ynf}).
The relevant matrix elements are in this case
\begin{align}
    &\bra{K^+} (ds)_R u_R \ket{p} \quad \mathrm{for}\quad Y(\phi)=2/3,\\
    &\bra{K^+} (du)_R s_R \ket{p} \quad \mathrm{for}\quad Y(\phi)=-1/3\,\mathrm{(}\lambda'\,\mathrm{coupling)},\\
    &\bra{K^+} (du)_L s_R \ket{p} \quad \mathrm{for}\quad Y(\phi)=-1/3\,\mathrm{(}\lambda''\,\mathrm{coupling)},
\end{align}
corresponding to $W_0 = 0.098\,\mathrm{GeV}^2$, $W_0 = -0.139\,\mathrm{GeV}^2$,  and $W_0 = 0.134\,\mathrm{GeV}^2$ respectively~\cite{Aoki:2017puj}.
The Wilson coefficient is the same for the three cases,
\begin{equation}
    C = \frac{\lambda \lambda'}{m_\phi^2} \left( \frac{m_\nu}{m_N} \right)^{1/2}.
\end{equation}
For the hypercharge $2/3$ mediator, this results in a partial lifetime
\begin{equation}
    \tau(p \rightarrow K^+ \nu) = 5.9 \times 10^{33}\, \mathrm{yr}\, \left( \frac{9\times10^{-19}}{\lambda'\lambda_2} \right)^2 \left( \frac{m_\phi}{1\,\mathrm{TeV}} \right)^4 \left( \frac{m_{N_2}}{0.5\,\mathrm{TeV}} \right) \left( \frac{0.05\,\mathrm{eV}}{m_{\nu_2}} \right),\label{eq:p_decay_tree_23}
\end{equation}
which we have normalized to the current limit~\cite{Workman:2022ynf} $\tau(p \rightarrow K^+ \nu) \geq 5.9\times 10^{33}\,\mathrm{yr}$.
The rate is a factor of $2$ shorter for the hypercharge $1/3$ mediator.
For the parameters of interest, the predicted lifetime in \cref{eq:p_decay_tree_23} is much below the experimental constraints, except in specific resonantly-enhanced baryogenesis scenarios.

\paragraph{Top-philic couplings.}

For the top-philic scenario, the proton decay rate depends on the hypercharge of the colored scalar mediator.
For the hypercharge $2/3$ mediator, assuming that the only nonzero couplings are $\lambda_{tN}$ and $\lambda'_{ds}$, the leading process is one loop as can be seen in the middle left diagram of \cref{fig:pdecay}.
For other flavor variations in the $\lambda'$ coupling, $\lambda'_{bd}$ and $\lambda'_{bs}$, similar diagrams with comparable rates are possible.
The Wilson coefficient for this diagram can be written as
\begin{align}
    C^{\phi_{2/3}}_{\rm top-ph} = \frac{1}{128\pi^2} \frac{g^2}{m_\phi^2} \left( \frac{m_\nu}{m_N} \right)^{1/2} V_{ud} V_{td}^* \, \lambda_{tN}\lambda'_{ds} \, \sqrt{m_t m_s} \Big( &I_2(x_t, x_s, x_W, 1) \nonumber \\ 
    &+\frac{1}{4m_W} I_4(x_t, x_s, x_W, 1) \Big),
\end{align}
where $x_f = m_f^2/m_\phi^2$ and the loop integral is
\begin{equation}
    I_n(x,y,w,z) = \int_0^\infty \frac{t^{n/2} \, \mathrm{d}t}{(t+x)(t+y)(t+w)(t+z)}.
\end{equation}
Using the relevant matrix element $\bra{\pi^+} (ud)_L d_R \ket{p}$, for which $W_0 = 0.186\,\mathrm{GeV}^2$~\cite{Aoki:2017puj}, we find
\begin{equation}
    \tau(p \rightarrow \pi^+ \nu) = 3.9 \times 10^{32}\, \mathrm{yr}\, \left( \frac{10^{-9}}{\lambda_{t{N_2}}\lambda_{ds}'} \right)^2 \left( \frac{m_\phi}{1\,\mathrm{TeV}} \right)^4 \left( \frac{m_{N_2}}{0.5\,\mathrm{TeV}} \right) \left( \frac{0.05\,\mathrm{eV}}{m_{\nu_2}} \right),\label{eq:p_decay_loop_23}
\end{equation}
which for the range of parameters of interest for baryogenesis is much shorter than the experimental constraint~\cite{Workman:2022ynf} $\tau(p \rightarrow \pi^+ \nu) \geq 3.9\times 10^{32}\,\mathrm{yr}$.
The results are similar for the other flavor combinations in the $\lambda'$ couplings.

The situation is different for the hypercharge $-1/3$ mediator with only $\lambda_{bN_i}$ and right-handed $\lambda'_{tb}$ couplings (the left-handed $\lambda''$ will be discussed below).
As can be seen in the lower right diagram in \cref{fig:pdecay}, in this case extra flavor-changing vertices are necessary and the rate is consequently much more suppressed.
The suppression can be estimated to be
\begin{equation}
    C^{\phi_{-1/3}}_{\rm top-ph} \sim G_F \, \frac{V_{ub} V^*_{ud} \, m_p^3}{m_b} \, C^{\phi_{2/3}}_{\rm top-ph},
\end{equation}
which leads to the lifetime
\begin{equation}
    \tau(p \rightarrow \pi^+ \nu) = 3.9 \times 10^{32}\, \mathrm{yr}\, \left( \frac{10^{-1}}{\lambda_{b{N_2}}\lambda'_{tb}} \right)^2 \left( \frac{m_\phi}{1\,\mathrm{TeV}} \right)^4 \left( \frac{m_{N_2}}{0.5\,\mathrm{TeV}} \right) \left( \frac{0.05\,\mathrm{eV}}{m_{\nu_2}} \right),\label{eq:p_decay_loop_23}
\end{equation}
orders of magnitude larger than current limits for the relevant parameter values.
For the left-handed $\lambda''$ couplings, even if only $\lambda''_{tb}$ (defined in the weak basis) is nonzero, after EWSB and CKM mixing, couplings to all generations are induced,
\begin{equation}
    \mathcal{L}\supset - \lambda''_{tb} \, \phi^\dagger \, \bar{Q}_3 Q_3^c \rightarrow - \lambda''_{tb} \, \phi^\dagger \, (U_{u})^\dagger_{3i} (U_{d})^T_{3j} \, \bar{u}_i d^c_j,
\end{equation}
where $U_{u}^\dagger U_{d} = V_{\rm CKM}$ denote the rotation matrices for the $u$ and $d$ quarks, $V_{\rm CKM}$ being the CKM matrix.
Thus, in the mass basis, the top-philic interaction induces (Wolfenstein-suppressed) couplings to all quark generations.
Consequently, proton decay is not expected to be suppressed for left-handed couplings as is for the right-handed coupling $\lambda'_{tb}$.
Depending on the exact form of $U_u$ and $U_d$ (only $V_{\rm CKM}$ is determined experimentally), we find proton decay rates
\begin{equation}
    \tau(p \rightarrow \pi^+ \nu) \lesssim 3.9 \times 10^{32}\, \mathrm{yr}\, \left( \frac{2\times 10^{-8}}{\lambda_{b{N_2}}\lambda''_{tb}} \right)^2 \left( \frac{m_\phi}{1\,\mathrm{TeV}} \right)^4 \left( \frac{m_{N_2}}{0.5\,\mathrm{TeV}} \right) \left( \frac{0.05\,\mathrm{eV}}{m_{\nu_2}} \right),\label{eq:p_decay_loop_23_2}
\end{equation}
which will end up enforcing a mild hierarchy $\lambda'' \ll \lambda'$. This may be explained by a supersymmetric origin of this twin Higgs theory, as discussed in \cref{sec:model_defn}.

SM dynamics do not require 
any rotations between the flavor- and mass-basis of right-handed quarks. It is therefore sensible to assume that the  $\lambda, \lambda'$ couplings, presumably originating in the flavor basis, are unchanged in the mass basis and thus retain their aligned top-philic character. 
That said, 
the underlying dynamics generating the SM quark flavor hierarchies may single out a preferred right-handed quark basis that is not aligned with the basis in which the $\lambda, \lambda'$ couplings are top-philic. 
Such misalignments could spoil the suppression of non-third generation couplings needed to avoid proton decay.
To quantify this effect, we parametrize any rotation between the flavor basis in which the $\lambda, \lambda'$ couplings are top-philic and the right-handed quark mass basis by the mixing angles $\theta_{bs}$, $\theta_{bd}$, $\theta_{tc}$, and $\theta_{tu}$.\footnote{If $\lambda''$ couplings are only generated radiatively through loop diagrams involving right-handed quarks and the Higgs, as would be the case in a supersymmetric UV completion, then they are are flavor-aligned in the same way the $\lambda'$ couplings and do not significantly contribute to the proton decay rate.}
Turning on the mixings a single one at a time, we find that avoiding the experimental bounds on proton decay requires
\begin{align}
\label{eq:flavormixing1}
    |\theta_{bs}| \lesssim 8 \times 10^{-4} \left( \frac{10^{-4}}{\lambda_{bN_2}\lambda'_{tb}} \right)^{1/2} \left( \frac{m_\phi}{1\,\mathrm{TeV}} \right) \left( \frac{m_{N_2}}{0.5\,\mathrm{TeV}} \right)^{1/4} \left( \frac{0.05\,\mathrm{eV}}{m_{\nu_2}} \right)^{1/4},\\
    |\theta_{bd}| \lesssim 6 \times 10^{-3} \left( \frac{10^{-4}}{\lambda_{bN_2}\lambda'_{tb}} \right)^{1/2} \left( \frac{m_\phi}{1\,\mathrm{TeV}} \right) \left( \frac{m_{N_2}}{0.5\,\mathrm{TeV}} \right)^{1/4} \left( \frac{0.05\,\mathrm{eV}}{m_{\nu_2}} \right)^{1/4},
\end{align}
while $\theta_{tc}$ and $\theta_{tu}$ remain unconstrained.
If all the mixing angles are turned on and are of similar size $\theta$, then it is necessary to require

\begin{equation}
\label{eq:flavormixing2}
    |\theta| \lesssim 2 \times 10^{-5} \left( \frac{10^{-4}}{\lambda_{bN_2}\lambda'_{tb}} \right)^{1/3} \left( \frac{m_\phi}{1\,\mathrm{TeV}} \right)^{2/3} \left( \frac{m_{N_2}}{0.5\,\mathrm{TeV}} \right)^{1/6} \left( \frac{0.05\,\mathrm{eV}}{m_{\nu_2}} \right)^{1/6} ,
\end{equation}
where the different coupling scaling arises due to multiple $\theta$ factors contributing to proton decay processes. 
These bounds constrain any UV flavor dynamics that are compatible with our baryogenesis scenario.

\subsection{Flavour Alignment}\label{sec:flavor_alignment}

As explained below \cref{e.MNpm}, having a single dominant reheaton much heavier than a GeV in the $\nu$MTH requires the $y_{ij}$ Yukawa couplings to be diagonal in the left- and right-handed neutrino bases that is most closely aligned with the mass basis before active-sterile mixing, with $y_1 \ll y_2, y_3$.
Furthermore, our analysis in \cref{sec:baryogenesis} will show that 
Baryogenesis additionally requires $\lambda_1 \ll \lambda_2, \lambda_3$ in the $\nu\phi$MTH. We now discuss how these alignments can arise in a manner compatible with the anarchic nature of the active neutrino mixing matrix $U_\mathrm{PMNS}$~\cite{ParticleDataGroup:2022pth}.

The issue can be demonstrated by considering how active neutrino masses and mixing are generated in the standard 3-generation type-1 see-saw mechanism, ignoring for the moment the B-sector and $\Delta M_{AB}$ mixing term between right-handed neutrinos in both sectors. 
The full $6 \times 6$ mass matrix is, in block form,
\begin{equation}
    M = \left( \begin{array}{cc}
    0 & m_{ij} 
    \\
    m_{ji} & M_{N_{ij}}
    \end{array}
    \right)
\end{equation}
where $i,j = 1,2,3$ run over active and sterile neutrino generations, $m_{ij} = y_{ij} v/\sqrt{2}$ and $M_N \gg m$. 
We take $M$ to be real for simplicity, which consistent with our assumption of real $y$ couplings.
Up to $\mathcal{O}(m^2/M^2)$ corrections, this can be brought into the block-diagonal form~\cite{Drewes:2013gca}
\begin{equation}
    M' = \left( \begin{array}{cc}
    m_{\nu_{ij}} & 0 
    \\
    0 & M_{N_{ij}}
    \end{array}
    \right) \ ,
\end{equation}
where active-sterile mixing is of order $m M^{-1}$, and 
the symmetric active neutrino mass matrix $m_{\nu} = - m M_N^{-1} m^T = U_\nu m_{\nu, diag} U_\nu^\dagger$ can be diagonalized with a unitary rotation $U_\nu$.
This illustrates the issue we have to address. The PMNS matrix is $U_{\rm PMNS} = U_\nu U_e^\dagger$, where $U_e$ are the rotations between the left-handed  charged   lepton flavor and mass basis. Since the charged lepton masses are highly hierarchical, $U_e$ should be close to $\mathds{1}$ in the absence of nontrivial cancellations. Therefore $U_{\rm PMNS} \approx U_\nu$, meaning $U_\nu$ has to have large mixing angles to reproduce the observed neutrino oscillation patterns. 
Importantly, the $N_i$ mass basis is determined by diagonalizing the $M_{N_{ij}}$ matrix. The $y_{ij} H \bar L_i N_j$ coupling is therefore generally not diagonal in the mass basis, and each $N_i$ mass eigenstate generally couples to all three neutrino mass eigenstates. It is therefore highly non-generic that one of the $N_i$ has a much smaller effective Yukawa coupling to its $\nu_L$ combination than the other two right handed neutrinos, but this is exactly the hierarchy required for the reheaton to be very long lived in the early universe. Furthermore, any significant rotations amongst the $N_i$ generically remove any hierarchies amongst the $\lambda_{mi}\, \phi\, q^c_m \bar N_i$ couplings, but baryogenesis and asymmetric reheating require $\lambda_1 \ll \lambda_2$ in the mass basis. 

In general, highly hierarchical coupling matrices in flavor space can be  generated by various UV frameworks, see~\cite{Altmannshofer:2022aml} for a modern review. However, the resulting matrices are very close to diagonal, and additional mechanisms must come into play to introduce significant misalignments. Given the various coupling hierarchies required for our model, and their required compatibility with large active neutrino mixing due to the right-handed neutrino nature of the reheaton, we now present one possible underlying flavor structure that could underpin our model and point towards possible UV symmetries and mechanisms that generate flavor in this scenario.

We start by imposing a $U(3)_{N_A} \otimes U(3)_{N_B}$ flavor symmetry on the right-handed neutrinos in the A and B sectors, which is only broken by the tiny non-universal $y, \lambda$ couplings.\footnote{
We note that large mixing in the sterile neutrino sector is not only incompatible with the $\lambda$ hierarchy, it also cannot by itself generate large mixing angles in $m_{\nu,ij}$ if $y_{ij}$ is highly hierarchical but close to diagonal. This, together with the resonance requirement for baryogenesis, further motivates this choice.
} 
This ensures that $M_{N_{ij}} = M_N \mathds{1}$ and $M_{AB} \propto \mathds{1}$ at tree-level.
Ignoring loop effects for the moment, we then assume that some UV flavor dynamics generates 
\begin{equation}
    y_{ij} \approx \left( \begin{array}{ccc}
    y_1 & & \\ & y_2 & \\ & & y_3 \end{array}\right)
    \ \ , \ \ 
    \lambda_i = (\lambda_1, \lambda_2, \lambda_3)
\end{equation}
with $y_1 \ll y_2 \sim y_3$ and $\lambda_1 \ll \lambda_2, \lambda_3$, and the hierarchical $y_{ij}$ matrix is assumed to be near-diagonal. 
We must assume that the effective basis of $(L, e)$ where the hierarchical and near-diagonal $y_e$ charged lepton Yukawa coupling is generated, is misaligned with the effective $(L, N)$ basis in which the above $y, \lambda$ coupling are generated.
One mechanism to achieve exactly this separation of hierarchy and mixing has been proposed in~\cite{Knapen:2015hia}.

In the $SU(2)_L$-preserving charged lepton mass basis, the right-handed neutrino Yukawa coupling is therefore $U_\nu y$, where $U_\nu \approx U_{\rm PMNS}$. This does not affect the RHNs or the $\lambda$ couplings, and after rotating the active neutrinos into their mass eigenbasis, the electroweak gauge interactions contain the $U_{\rm PMNS}$ matrix and the Yukawa coupling takes the form assumed in our analysis: $\frac{y_i}{\sqrt{2}} (h+v) \bar \nu_{L_i} N_i$, with $y_1 \ll y_2, y_3$ and each active neutrino mass eigenstate $\nu_{L_i}$ having $m_{\nu_i} = y_i v/(\sqrt{2} M_N)$.

This is not only the exact scenario that we assume for our analysis, it has the further advantage that it  automatically realizes the resonant baryogenesis requirement of the viable top-philic scenario in \cref{s.topphilic}. The right-handed neutrino mass matrix receives loop corrections of order 
\begin{equation}
    M_{N_{ij}} = M_N\left[1 + \mathcal{O}\left( \frac{y_i y_j}{16 \pi}\right) + \mathcal{O}\left( \frac{\lambda_i \lambda_j}{16 \pi}\right)\right] \, .
\end{equation}
This does not spoil the suppressed decay of $N_1$ required for asymmetric reheating, nor does it disrupt the $\lambda$ coupling hierarchies, but it does shift the mass of $N_2$ by the exact amount needed for maximum resonance enhancement $\Delta M \approx \Gamma_{N_2}/2$ of twin- and SM baryogenesis in the top-philic $\nu\phi$MTH model, as will be discussed in \cref{sec:baryogenesis}.

\subsection{Cosmological History for Baryogenesis}

In the $\nu\phi$MTH, small values of the $\lambda$ couplings allow a small fraction of $N_i$'s to decay to visible and  twin quarks. If that process occurs out-of-equilibrium, and there is interference between a loop- and tree-level decay process, then the decays of the RHN can generate a baryon asymmetry. We compute predictions for these asymmetries in the next section, but before doing so, it is useful to specify the kind of cosmological history for which this model can realize both asymmetric reheating and baryogenesis. 

In order for the decay of a RHN $N_i$ to generate any asymmetry due to the interference of a tree- and loop-induced decay process to quarks, the loop  must include particles with mass below that of $N_i$. Specifically, this loop can involve $h_A, H_B$ for decays to either sector, or $\phi_{A(B)}$ for decays to the visible (twin) sector. Since new colored states below the Higgs mass are excluded, baryogenesis in the visible sector requires $M_{N_i} > m_h$ for the RHN that generates the asymmetry. 

One might ask if the $N_i$ that generates the asymmetry has to be the same $N_j$ that dominates the asymmetric reheating. If they are different mass eigenstates, then both have to decay out-of-equilibrium, though only the reheaton has to completely dominate the energy density of the universe. However, since the asymmetry-generating $N_i$ has to be heavier than 125 GeV, making it decay even modestly out-of-equilibrium
requires such a small Yukawa coupling that the corresponding active neutrino mass $m_{\nu_i}$  is much smaller than 0.1 eV. Therefore, the other $N_j$ have to have larger Yukawas to generate the $m_{\nu_j} \sim 0.1~$eV required to satisfy neutrino oscillation bounds~\cite{Esteban:2020cvm}, which precludes them from being long-lived enough to sufficiently suppress $\Delta N_\mathrm{eff}$, see \cref{fig:dneff}.

We can therefore conclude that if we want the $\nu\phi$MTH model to accommodate cosmological bounds and account for the matter-antimatter asymmetry of the universe, then both the asymmetric reheating and the baryon asymmetries must be generated in the decay of the same RHN state $N_1$ with a mass satisfing $M_{N_1} > 125$~GeV.

\section{Baryon Asymmetry and Prediction for Atomic Dark Matter}
\label{sec:baryogenesis}

In order to generate any matter-antimatter asymmetry, the Sakharov conditions~\cite{sakharov} must be satisfied. 
By design, $N_1$ decays out of equilibrium at the time of asymmetric reheating, and the structure of its couplings violates baryon number. 
Therefore, only the size of charge-parity violation (CPV) in $N_1$ decays remains to be calculated. 
CPV requires at least two RHN  generations with non-aligned phases in their couplings.
We thus include a second $N_2$ with mass $m_{N_2}$ in our calculations in this section.\footnote{Including $N_3$ would not significantly change our results, except for a modest $\mathcal{O}(1)$ enhancement in the generated asymmetries for regions of parameter space where both $N_2$ and $N_3$ have comparable effects.} We regard $m_{N_2}$ as a free parameter in the analyses of this Section, but our results, especially in the top-philic scenario, will bear out the universal right-handed-neutrino mass scenario discussed in \cref{sec:flavor_alignment}.

The Yukawa couplings of $N_{1,2}$ to the SM leptons are approximately fixed: $y_1$ is set by the reheating temperature $T_{A,R}$, see 
\cref{fig:dneff}, and $y_2$ is set by requiring the mass of the second-lightest neutrino to be $m_{\nu_2}\sim 0.1 $ eV, which leads to $ y_2 =\sqrt{m_{\nu_2}M_{N_2}/v} \sim 10^{-6} \sqrt{M_{N_2}/(500~\mathrm{GeV})}$.
We also require that the $N_1$ decays which generate the baryon asymmetry do not significantly modify the asymmetric reheating mechanism:\footnote{One could imagine scenarios where the reheating asymmetry and the baryon asymmetry are both generated by $N_1$ decays to $\phi_{A,B}^{(*)}$, but this would not only restrict us to generating much fewer twin baryons than visible baryons, it would place many additional restrictions on the model. The complicated hybrid case where the first row of \cref{e.ARHdominant} is satisfied but not the second would  correspond to larger $\dneff$ than we assume in our calculation.}
\begin{equation}
\label{e.ARHdominant}
\begin{array}{lllll}
    \Gamma_{N_1} &\approx& \Gamma(N_1 \to \mbox{A-sector})) &\gg& \Gamma(N_1 \to \phi_{A}^{(*)} + \ldots),
    \\
    \Gamma_{N_1} &\gg& \Gamma(N_1 \to \mbox{B-sector})) &\gg& \Gamma(N_1 \to \phi_{B}^{(*)} + \ldots).
\end{array}
\end{equation}
In this calculation it will also be crucial to separately account for the $\mathbb{Z}_2$-even and -odd $N_{1,2\pm}$  mass eigenstates, since $B$-sector couplings of the $N_{1,2-}$ carry an extra minus sign, see \cref{e.LY23} and \cref{e.LY13}. 
Depending on the choice of $\phi_{A,B}$ couplings to quarks (generation universal vs.~top-philic), as well as the  masses of $N_i$ and $\phi_{A,B}$, different decays dominate the asymmetry generation. 

Since we do not introduce any explicit $\mathbb{Z}_2$ breaking, we assume all couplings to be the same in both sectors. For concreteness, we also assume that the $y$ couplings are real so that complex phases arise from $\lambda$-type couplings alone.
As we will see, this means that some coupling combinations in the twin sector can be set by requiring generation of the known baryon asymmetry in the visible sector~\cite{Planck:2018vyg},
\begin{gather}\label{eq:eta}
    \eta = \frac{n_b-n_{\Bar{b}}}{s} = \frac{n_N}{s}\, \epsilon_{CPV}
    \approx 
    (6.10 \pm 0.4) \times 10^{-10}  \ ,
\end{gather}
where we assume the central value for simplicity.

The CPV parameter $\epsilon_{CPV}$ is calculated as an interference between tree and 1-loop level diagrams in the decay of $N_1$. Generically, the matrix element of the decay can be writen as
\begin{gather}
\label{eq:c0c1}
    \mathcal{M} = c_0 \mathcal{A}_0 + c_1 \mathcal{A}_1,
\end{gather}
where $c_0$ and $c_1$ are the coupling combinations setting the size of the relevant tree level and 1-loop diagrams respectively, and $\mathcal{A}_0$ and $\mathcal{A}_1$ are the tree level and 1-loop level amplitudes. With this parametrization, $\epsilon_{CPV}$ is given by
\begin{gather}
    \label{eq:eta}
    \epsilon_{CPV} = \frac{\Gamma(N\rightarrow q+X)- \Gamma(N\rightarrow\bar{q} + X)}{\Gamma_\text{tot}} = \Im(c_0c_1^*)\, \frac{\sum_\alpha\int 2 \mathcal{A}_0 \Im(\overline{\mathcal{A}}_1)d \Pi_{LIPS}}{\Gamma_{N_1}^{\text{tot}}}  \, ,
\end{gather}
where $X$ are other possible final states in the decay (with implied dagger or bar in the second term as appropriate).
The sum over decay final states $\alpha$ is different for the generation universal and top-philic scenarios. 
In the former case we sum over all  generation of quarks, whereas in the latter only quarks with sizeable $\lambda, \lambda', \lambda''$ couplings (mostly third generation) are included. With this, the abundance of atomic Dark Matter can be computed as 
\begin{gather}
    \Omega_{aDM} = \Omega_b \frac{m_{p_B}}{m_{p_A}} \frac{\eta_B}{\eta_A},
\end{gather}
where the ratio of proton masses is given by \cref{e.mpBovermpA}, and $\Omega_b\approx 0.03$ is abundance of baryons in the SM sector. 
We now calculate predictions for the aDM fraction $\hat r  = \Omega_{aDM}/\Omega_{DM}$ in the different possible $\nu\phi$MTH scenarios.

\subsection{Generation Universal Couplings}
\label{s.genuniversal}
In the generation-universal couplings scenario there are two possibilities for $N$-decays into SM or twin quarks. 
Either $M_N> m_{\phi_{A,B}}+m_{p_{A,B}}$ and the decay of $N$ is dominantly two body, or $m_{\phi_{A,B}} +m_{p_{A,B}} > M_N$ and decays are dominantly three body. 
To streamline the discussion we present the results for the $Y_\phi=2/3$ mediator, since the conclusions only change minimally for the case of $Y_\phi =  -1/3$.

\begin{figure}
    \centering

    \begin{tabular}{c}
  \includegraphics[scale=1]{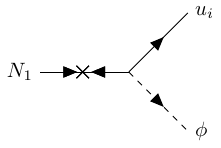}\,%
  \\  
  \includegraphics[scale=1]{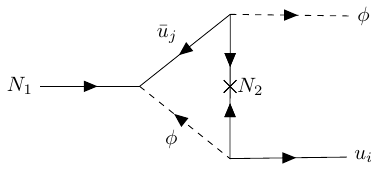}\,%

    \includegraphics[scale=1]{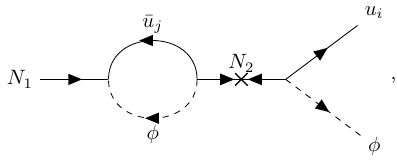}\,%
    \\    
  \includegraphics[scale=1]{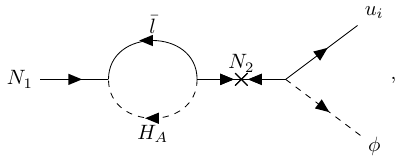}\,%

    \end{tabular}
    
    (a) \vspace{6mm}
    
    \begin{tabular}{c}
    
    \includegraphics[scale=1]{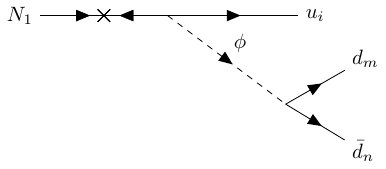}\,
    \\
     \includegraphics{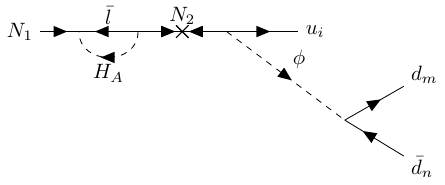}\,%
    \end{tabular}
    
        (b)
    \caption{
    Diagrams that generate CPV for generation universal couplings in either the $A$ or $B$ sector.
    CPV occurs due to interference between tree level diagram with one loop diagrams. $u$ and $d$ stand for up-type and down-type quarks, respectively, and $i,j,m,n$ are flavor indices.
    (a) Two-body reheaton decay $M_N>m_\phi$. 
    If $\lambda_1\lambda_2>y_1y_2$, the dominant CPV effect is from interference of the first row diagrams with the second row. 
    If $y_1y_2>\lambda_1\lambda_2$ then dominant CPV effect is from interference of the first row diagrams with the third row.
    (b) Three-body reheaton decay $M_N<m_\phi$.
    CPV is only generated from interference between the first and second row. (There are also diagrams with $\phi$ in the loop, but as we explain in the text, those do not contribute in our case.)
    The arrows indicate the flow of two-component chirality (for fermions) and scalar charge for scalars. 
    Note that versions of these loop diagrams with the mass insertion on the $N_1$ leg do not contribute to $\epsilon_{CPV}$. 
    In the resonant case, the triangle diagram is subdominant since it is not enhanced.}
    \label{fig:diagrams_2}
\end{figure}

Let us focus on the SM sector first. Consider the case when $M_{N_1} > m_{\phi_A} +m_{p_{A}}$.
The dominant CP violating decays are shown in the second and third rows of \cref{fig:diagrams_2} (a). 
The dominant CPV effects are generated by the second row diagrams with $\phi_A$ in the loop in the case $\lambda_1\lambda_2 > y_1y_2$, otherwise it is generated by the third row ones with $H_A$ in the loop.
Now, consider the case when $M_{N_1} < m_{\phi_A}+m_{p_{A}}$. The dominant Feynman diagrams are shown in \cref{fig:diagrams_2} (b). In this case we have the SM Higgs $H_A$ and an A-sector lepton running in the loop. 
In either scenario, the optical theorem mandates that the Higgs boson (or the heavier $\phi_A$) and the neutrino (quark) be on-shell 
in order for $\Im(\mathcal{A}_1)\neq 0$, and this leads to the lower bound on the RHN mass $M_{N_1}\gtrsim 125$~GeV.

Using \cref{eq:eta} we can constrain the coupling combination needed for this model to generate baryogenesis in the SM sector. 
First, note that  $\epsilon^{A,B}_{\text{CPV}}$ factorizes as
\begin{gather}\label{eq:coupcombo}
    \epsilon^{A,B}_{\text{CPV}} = \sum_q \epsilon^{A,B}_{\text{CPV,}q} = \Im\left[\delta_{A,B}(\lambda_1,\lambda_2,\lambda',y_1,y_2)\right] \sum_q \mathcal{F}(M_{N_1},M_{N_2},m_{\phi_{A,B}},m_{q_{A,B}},T_{(A,B),R}),
\end{gather}
where $\mathcal{F}$ is the ratio of loop functions and the total $N_1$ decay width, which depends on  $T_{A,R}$ since this sets $y_1$ and hence $\Gamma_{N_1}$.
$\delta_A$ is a combination of the $\lambda, \lambda', y$ couplings:
\begin{gather}
\delta_{A}(\lambda_1,\lambda_2,\lambda') = 
\left\{ \begin{array}{ll} 
\lambda_1^{*2}\lambda_2^2 & \text{if $N_1$ decay is 2-body and $\lambda_1\lambda_2 > y_1y_2$}
\\
y_1y_2\lambda^*_1\lambda_2  & \text{if $N_1$ decay is 2-body and $y_1y_2 >\lambda_1 \lambda_2$}
\\
y_1y_2\lambda^*_1\lambda_2\lambda'^2 &  \text{if $N_1$ decay is 3-body}
\end{array} 
\right. \ .
\end{gather} 
The observed matter antimatter asymmetry in the visible sector together with \cref{eq:eta} can then be used to determine value of $\delta_A$.

This makes clear why perturbativity and proton decay constraints make this scenario generically unviable. 
In the 2-body case, $\phi_A$ has to be short-lived enough not to disrupt BBN. The decay width for $\phi_A$ is given by
\begin{equation}
    \Gamma_{\phi_A} =\frac{1}{\tau_{\phi_A}} \simeq \frac{9\lambda'^2}{16 \pi} m_{\phi_A}.
\end{equation}
For $\tau_{\phi_A} < 0.1$~s and $m_{\phi_A} \sim 550$~GeV as allowed by the gap in current LHC exclusions, this implies $\lambda' \gtrsim 10^{-14}$. 
Proton decay constraints require $\lambda_{1,2} \lesssim 5\cdot 10^{-5}$, see \cref{eq:p_decay_tree_23}, while the asymmetric reheating bound on $\lambda_1$ from \cref{e.ARHdominant} is many orders of magnitude stronger. For such small couplings, $\delta_A$ is far too small to generate the required $\eta$ in the visible sector in \cref{eq:eta}. 
If the decay to SM quarks is 3-body, the generated baryon asymmetry is even smaller.

This conclusion can be circumvented in a resonant baryogenesis scenario, where the reheaton decay to SM quarks is 2-body, but the asymmetry is enhanced by considering $M_{N_2} \approx M_{N_1}$ (this is similar to the resonant leptogenesis scenario, see \cite{Pilaftsis:2003gt}). When the two RHN are close in mass, one has to take into account their small but nonzero decay width in the propagator. The triangle diagram in \cref{fig:diagrams_2} is irrelevant in this limit as it does not receive any resonant enhancement. 
The CP asymmetry becomes proportional to
\begin{gather}\label{eq:resonantcpv}
\epsilon_{\text{CPV,A}} \propto \frac{M_{N_1}^2(M_{N_1}^2-M_{N_2}^2)}{(M_{N_1}^2-M_{N_2}^2)^2 + M_{N_1}^2 \Gamma_{N_2}^2}\, \text{Br}(N_1\rightarrow q_A\phi_A),
\end{gather}
in which the decay width of $N_2$ is given by
\begin{gather}
    \Gamma_{N_2} \simeq \frac{M_{N_2}}{16 \pi} \left( y_2^2 + 9 \lambda_2^2 \left(2- \frac{m_{\phi_A}^2}{m_{N_2}^2}- \frac{m_{\phi_B}^2}{m_{N_2}^2}\right)\right).
\end{gather}

\begin{figure}    \centering
  \includegraphics[scale=.45]{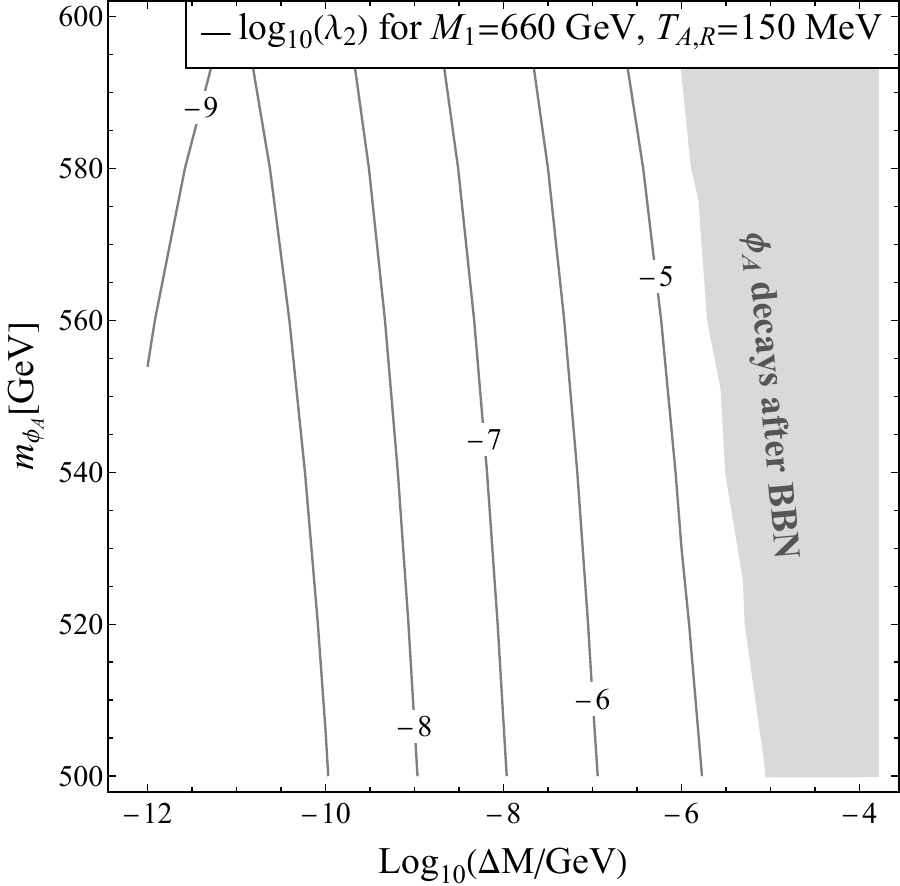}\,%
    \caption{
    Values of the $\lambda_2$ coupling 
    required to generate the visible baryon asymmetry through decays of the $N_1$ reheaton, 
    in the $\nu\phi$MTH model with $Y_\phi = 2/3$ and generation-universal $\phi$ couplings to quarks, in the regime of resonant-enhanced asymmetry with  $M_{N_1}\approx M_{N_2}$, for $M_{N_1}=660 $ GeV, $T_{A,R}=150 $ MeV. Here, $\lambda_1$ is set to $\lambda_1^\mathrm{max}$ to satisfy \cref{e.ARHdominant}. 
    The resonant enhancement allows for a sufficiently small $\lambda_{1,2}$ to satisfy proton decay constraints, see \cref{eq:p_decay_tree_23}, as well as the $\phi_A$ lifetime requirements from SM BBN. 
    The peak near $\Delta M =  M_{N_1}-M_{N_2} \sim 10^{-11}~\mathrm{GeV} \sim \Gamma_{N_2}/2$  corresponds to the maximal resonant enhancement.
    }
    \label{fig:gen_res}
\end{figure}

As a simplifying assumption, we set the phase of $\delta_A$ to be maximal ($=\pi/2$). With this, the required $\delta_A = y_1 y_2\lambda_1 \lambda_2$ coupling combination to produce the observed visible baryon asymmetry can be computed. The $y_1$ and $y_2$ couplings are set by the requirements of asymmetric reheating and by the mass of the second lightest neutrino ($m_{\nu_2}\approx 0.1$~eV), respectively. The maximum $\lambda_1^{\text{max}}$ is set by satisfying the inequalities in \cref{e.ARHdominant} by a factor of 10 in the widths.  By assuming $\lambda_1=\lambda_1^{\text{max}}$, we show contours of $\lambda_2$ in the $\Delta M = M_{N_1}-M_{N_2}$ vs.~$m_{\phi_A}$ plane in \cref{fig:gen_res}. The shaded region shows where $\lambda_2$ is too large to satisfy $\lambda_2 \lambda' \lesssim 9\cdot10^{-19}$ and $\lambda' \gtrsim 10^{-14}$, in which case the decay of $\phi_A$ occurs after BBN.
In this resonant regime, 
generating the required baryon asymmetry in the visible sector is easily compatible with the constraints on $\lambda_{1,2}$ coming from the $\phi_A$ lifetime and proton decay. Note that not disrupting the asymmetric reheating mechanism as quantified by \cref{e.ARHdominant} requires $\lambda_2 \gg \lambda_1$.
The maximum resonant enhancement is achieved for $ M_{N_1}-M_{N_2} \sim \Gamma_{N_2}/2 \sim 10^{-11}$ GeV, as is clearly visible in \cref{fig:gen_res}.

In the B sector the situation is slightly different. 
Consider first the case in which the reheaton decay to $\phi_B^*$ is three body. In that scenario, the asymmetry is generated by the interference of diagrams in \cref{fig:diagrams_2} (b), as well as similar diagrams with $\phi_A$ (but not the heavy $\phi_B$) in the loop. However, for both of these cases, diagrams with even and odd combinations of $N_{1,2\pm}$ participating in the process have different signs,
which leads to a large cancellation and suppresses the asymmetry by $M_{AB}/M_N \ll 10^{-3} - 10^{-2}$.
Therefore, this is a viable scenario for asymmetric reheating and baryogenesis in the visible sector, but the twin baryon asymmetry is essentially negligible.

\begin{figure}   
\centering
  \includegraphics[scale=.41]{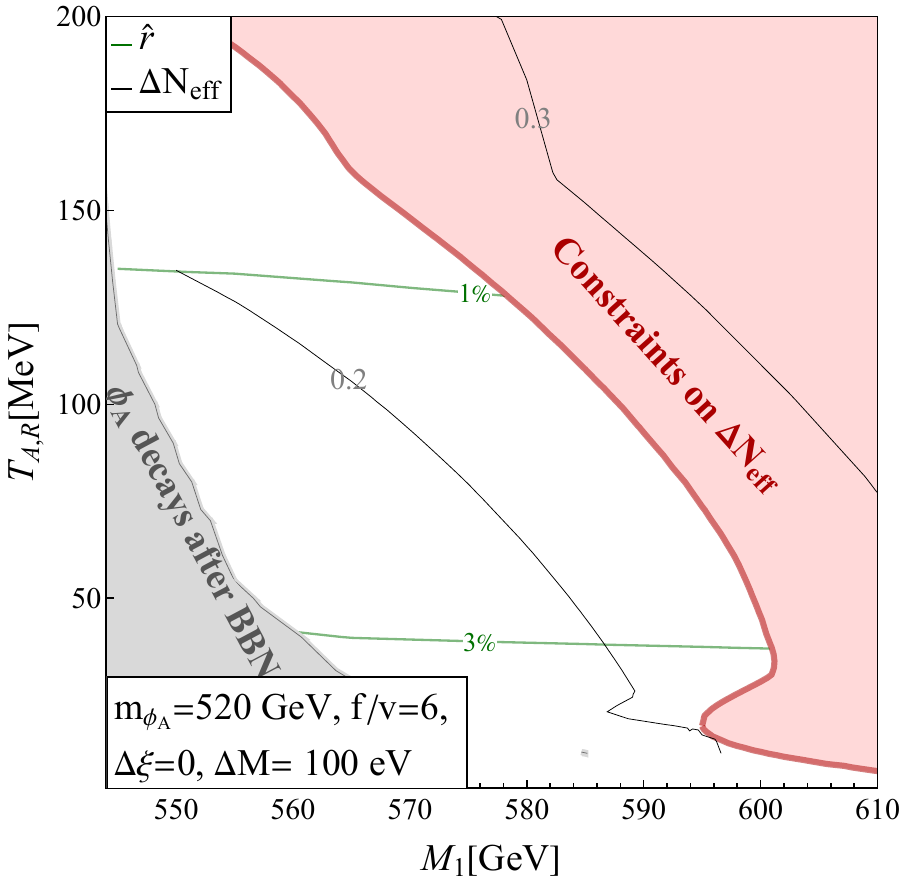}\,%
  \includegraphics[scale=.4]{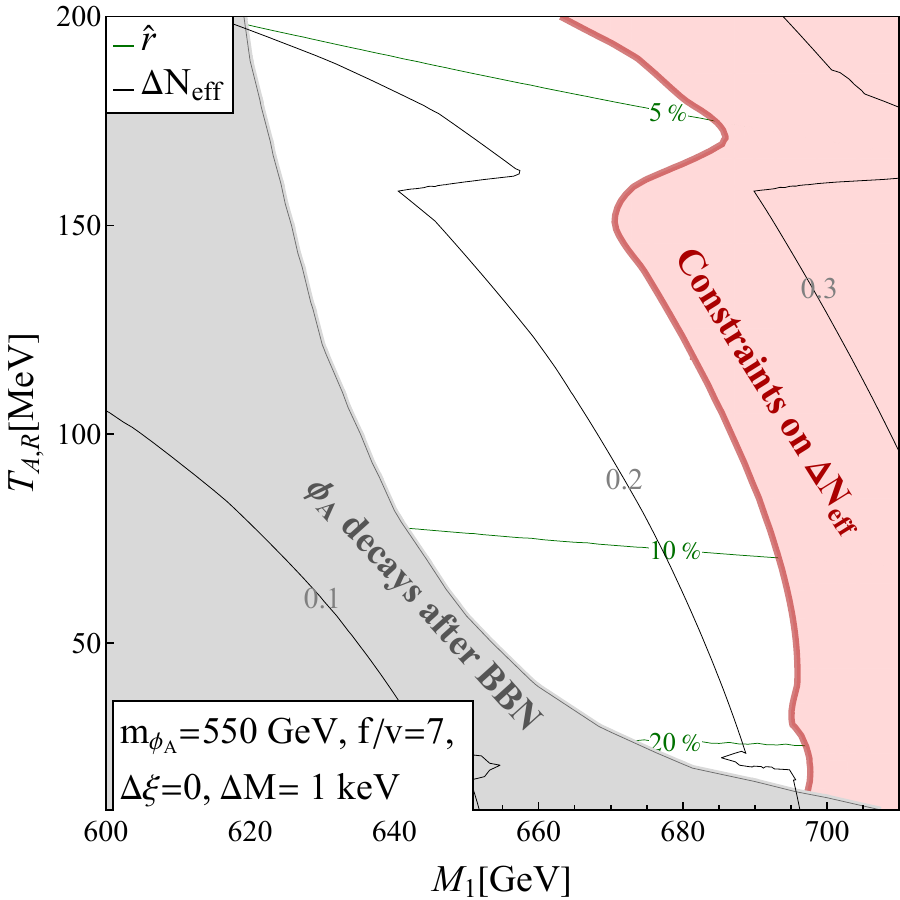}\,%
  
\includegraphics[scale=.4]{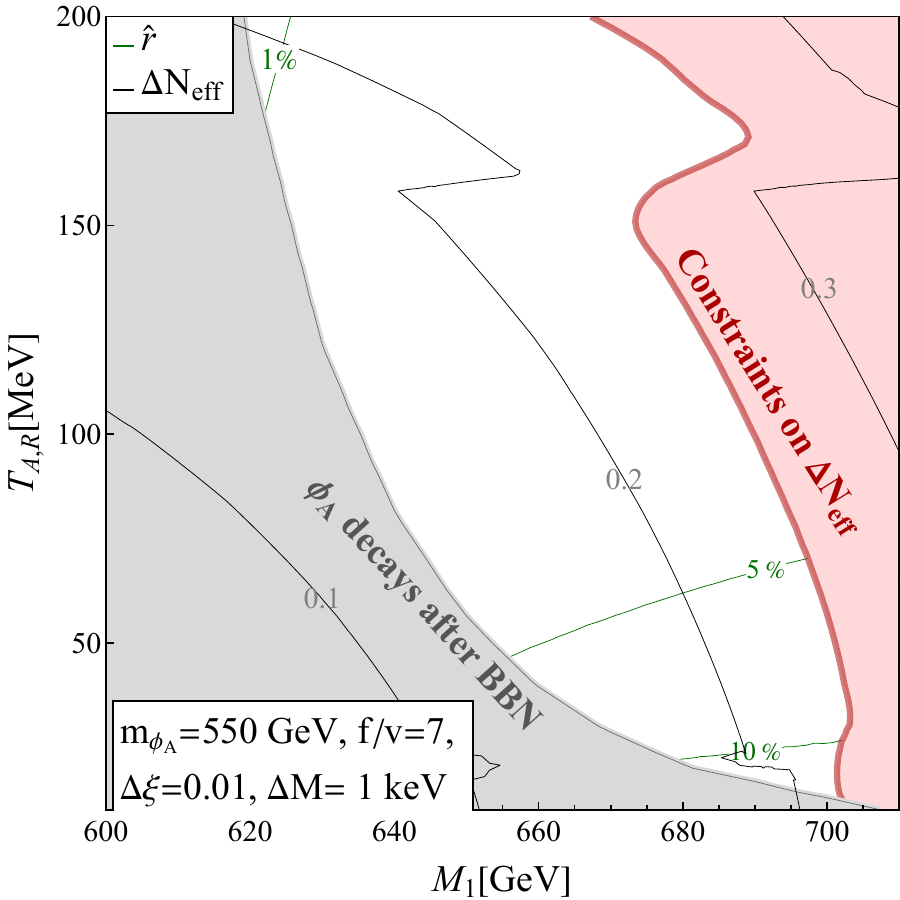}\,%
\includegraphics[scale=.4]{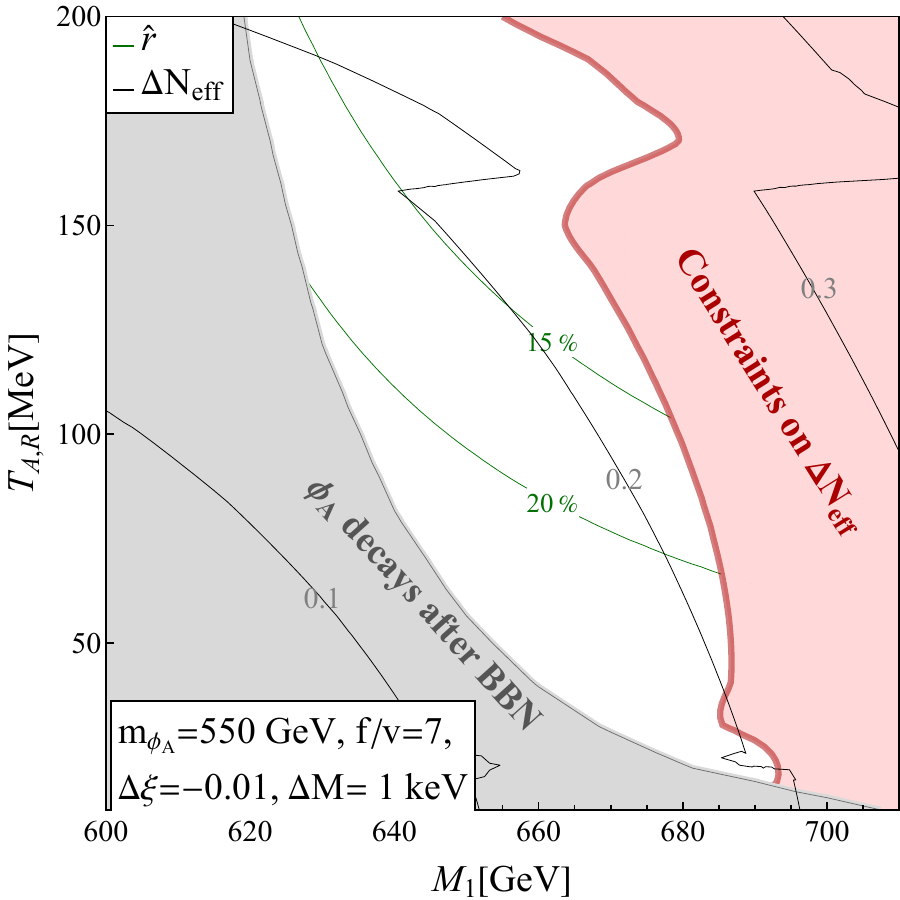}\,%

    \caption{Contours of aDM fraction for the resonant generation-universal couplings scenario (green), for different values of $f/v=6,7$ and $\Delta\xi=0,\pm0.01$. In the figures with $f/v=7$, $\Delta M= 1$ keV and $m_{\phi_A}=550$ GeV, and for $f/v=6$, $\Delta M = 100$~eV and $m_{\phi_A}=520 $~GeV in the gap of LHC exclusions~\cite{CMS:2022usq}. 
    We also show the prediction for $\dneff$ (black contours), the enhanced constraints on $\dneff$ that take the generated twin baryon abundance into account~\cite{Bansal:2021dfh} (red shading), and regions where $\phi_A$ decay disrupts BBN (gray shading).   
    Note that a wide range of $\Omega_{aDM}$ values can be generated, and that the twin baryon abundances scale with
    $M_{N_1}-M_{N_2}$.
    }
    \label{fig:results1}
\end{figure}

Let us then consider the case in which the reheaton decay to $\phi_B$ is two body as well. In that case, only the diagrams in the second row of \cref{fig:diagrams_2} (a) contribute to the twin asymmetry, since the diagrams in the third row are suppressed by $M_{AB}/M_N$ as above. 
We then have $\delta_A\simeq y_1y_2\lambda_1 \lambda_2$ but $\delta_B \simeq \lambda_1^{*2} \lambda_2^2$, and therefore the aDM abundance is given by
\begin{gather}
    \Omega_{aDM} = \Omega_b \frac{m_{p_B}}{m_{p_A}} \frac{\lambda_1\lambda_2}{y_1y_2} \frac{\Gamma(N_1\rightarrow q_B \phi_B)}{\Gamma(N_1\rightarrow q_A \phi_A)} \frac{\text{Im} (\mathcal{F}_{\phi_B  \mathrm{loop}})}{\text{Im} (\mathcal{F}_{H_A  \mathrm{loop}})}.
\end{gather}
Here, the phases $\delta_{\text{A,B}}$ are taken to be $\pi/2$. Taking into account that the decays in both sectors are 2-body, we conclude that $\Omega_{aDM} \approx  \Omega_b (m_{p_B}/m_{p_A}) \cdot (\lambda_1\lambda_2)/(y_1y_2)$ up to $\mathcal{O}(1)$ factors. Due to the requirement $\lambda_1\ll y_1$ from \cref{e.ARHdominant}, sitting exactly on the resonant peak would generate negligible abundances of atomic dark matter, since in that case $\lambda_2\ll y_2$.
Non-negligible twin baryon abundances require a $M_{N_2} - M_{N_1}$ mass difference larger than the minimal value generated by radiative corrections in the approximate $U(3)_N^2$-symmetric scenario of \cref{sec:flavor_alignment}.
For concreteness, we take $\Delta M = M_{N_1} - M_{N_2} \sim 10^{-6} \text{ GeV} = 1 $ keV in \cref{fig:gen_res}, and note that $\Omega_{aDM}$ scales with $M_{N_1}-M_{N_2}$ 
in this off-resonance regime.
\cref{fig:results1} then shows contours of the aDM fraction (green contours), together with the predicted values of $\dneff$ (black contours) and cosmological constraints that include the effect of the generated twin baryon abundances~\cite{Bansal:2021dfh} (red shading). The gray shaded region represents parts of parameter space where $\lambda_2$ is too large and late $\phi_A$ decay disrupts BBN. 
The dependence of our results on $f/v$ and different scalar masses $\Delta \xi \neq 0$ is shown in the top and bottom panels, respectively. 
As anticipated, the situation for the $Y_\phi = -1/3$ model is qualitatively similar.

We conclude that the generation-universal $\nu\phi$MTH model can generate a wide range of possible aDM fractions, from negligible to $O(1)$, as long as the colored scalar in the visible sector lies in a narrow mass window around $m_{\phi_A} \approx 550~\mathrm{GeV}$ not yet excluded by LHC searches~\cite{CMS:2022usq}. This also requires a resonant enhancement with $M_{N_2} \approx M_{N_1}$, but with a greater mass difference than the minimal radiative contribution from the different $y, \lambda$ couplings, as well as $\lambda_1 < y_1 \ll  y_2 \sim \lambda_2$ and the various alignments discussed in \cref{sec:flavor_alignment}. 
While this is a viable scenario, the new colored scalar below the TeV scale 
goes against the phenomenological motivation for the Twin Higgs mechanism, and may be completely excluded by direct LHC searches in the near future.
This motivates the top-philic coupling scenario we study below.

\subsection{Top-philic coupling scenario}
\label{s.topphilic}

Although the generation-universal scenario has a simple coupling structure, it requires the existence of a new light colored scalar below a TeV, an undesirable feature in a Neutral Naturalness model even if the mass of that new colored state is not directly related to tuning in the theory.
This can be circumvented by assuming the existence of some flavor hierarchy in the scalar mediator couplings.
In particular, a top-philic scenario generates much more suppressed contributions to proton decay than the generation-universal coupling scenario.
The observed visible baryon asymmetry 
can then be generated 
in reheaton decays involving an off-shell $\phi_A$ and sizeable $\lambda^{\prime(\prime)}$ couplings (which for this scenario are not constrained by the LHC, see \cref{sec:lhcconstraints}). 
Thus, the new colored scalars can have the high masses that are associated with a Twin Higgs UV completion and/or avoid LHC bounds.

In what follows, we focus on the $Y_\phi = -1/3$ scenario where the reheaton decay to $\phi_A^*$ is three-body, with $m_{\phi_A} > M_{N_1} - m_{t_A}$, and the only relevant scalar couplings in both sectors are  $\lambda_{ib}\, N_i\, b\,\phi$ and $\lambda'_{tb}\, \phi\, t\, b$, simply referred to as $\lambda_i$ and $\lambda'$ in this section.
We set $\lambda'' = 0$ to make the calculations more transparent, but we discuss below how our calculations apply for the more general case with nonzero $\lambda''$. 
We do not discuss the $Y_\phi=2/3$ mediator, since scenarios with viable baryogenesis and cosmology are entirely excluded by proton decay constraints, see \cref{sec:pd}.

For the range of $f/v$ that we consider, the decays in B sector can either be 2-body ($N_1\rightarrow b_B \phi_B$), 
3-body ($N_1\rightarrow 2b_B+t_B$), 
4-body ($N_1\rightarrow 3b_B + W_B$) or 5 body ($N_1\rightarrow 3b_B + $ 2 light twin quarks). 
The 3- and 4-body decay scenarios are incompatible with asymmetric reheating, since they require $M_{N_1} > m_{W_B}$. 
On the other hand, the 5-body decay in the twin sector leads to negligible twin baryon asymmetry, not only due to the phase space suppression compared to the  3-body decay in the visible sector, but also because the dominant contribution to the asymmetry in both sectors comes from an $H_A$ loop, resulting in an additional $M_{AB}/M_{N_1}\lesssim 10^{-3} - 10^{-2}$ suppression in the $B$-sector only (see discussion in \cref{s.genuniversal}). 
We have quantitatively verified that this can be a viable scenario of asymmetric reheating and  baryogenesis, but one where no significant atomic dark matter abundance is generated.

\begin{figure}
    \centering
    \includegraphics[scale=1]{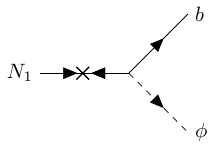}\,
    \includegraphics[scale=1]{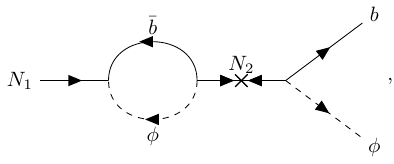}\,
    
    \includegraphics[scale=1]{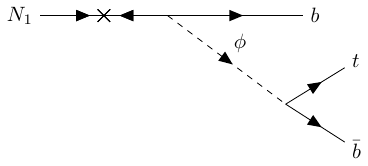}\,
    \includegraphics[scale=1]{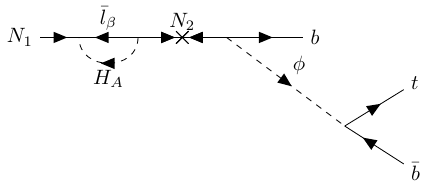}\,
    \caption{
    Diagrams relevant for CPV effects in the top-philic coupling scenario with $M_{N_2} \approx M_{N_1}$. In the B sector, $N_1 \to \phi_B$ is two-body (top row), and the tree-level diagram interferes with the $\phi_B$ loop. The corresponding diagram with $H_A$ in the loop is subdominant due to the $M_{AB}/M_N$ suppression in the resonant scenario. 
    In the A sector, $N_1 \to \phi_A^{(*)}$ is three-body (bottom row), and the dominant interference comes from the diagram with $H_A$ in the loop. The corresponding diagrams with $\phi_A$ ($\phi_B$) in the loop generate no (suppressed) asymmetry since $\phi_A$ is too heavy (the contribution is suppressed by $M_{AB}/M_N$).
    }
    \label{fig:topphilicdiag}
\end{figure}

\begin{figure}
    \centering
    \includegraphics[width=0.49\linewidth]{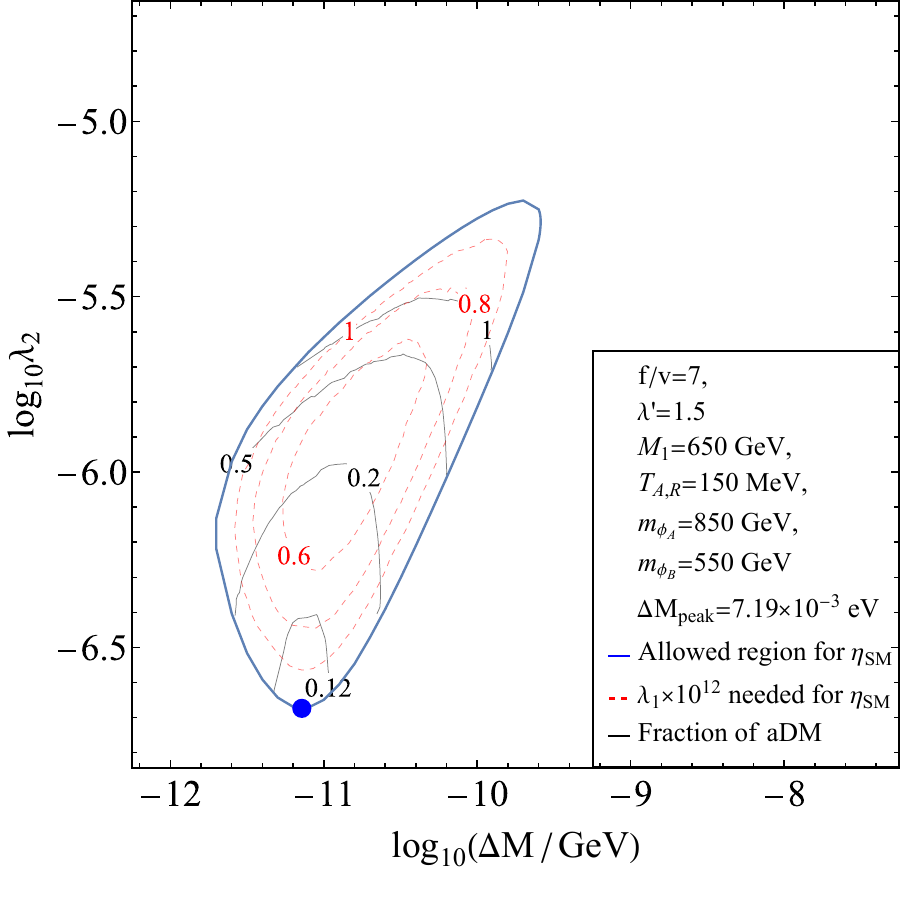}
    \includegraphics[width=0.49\linewidth]{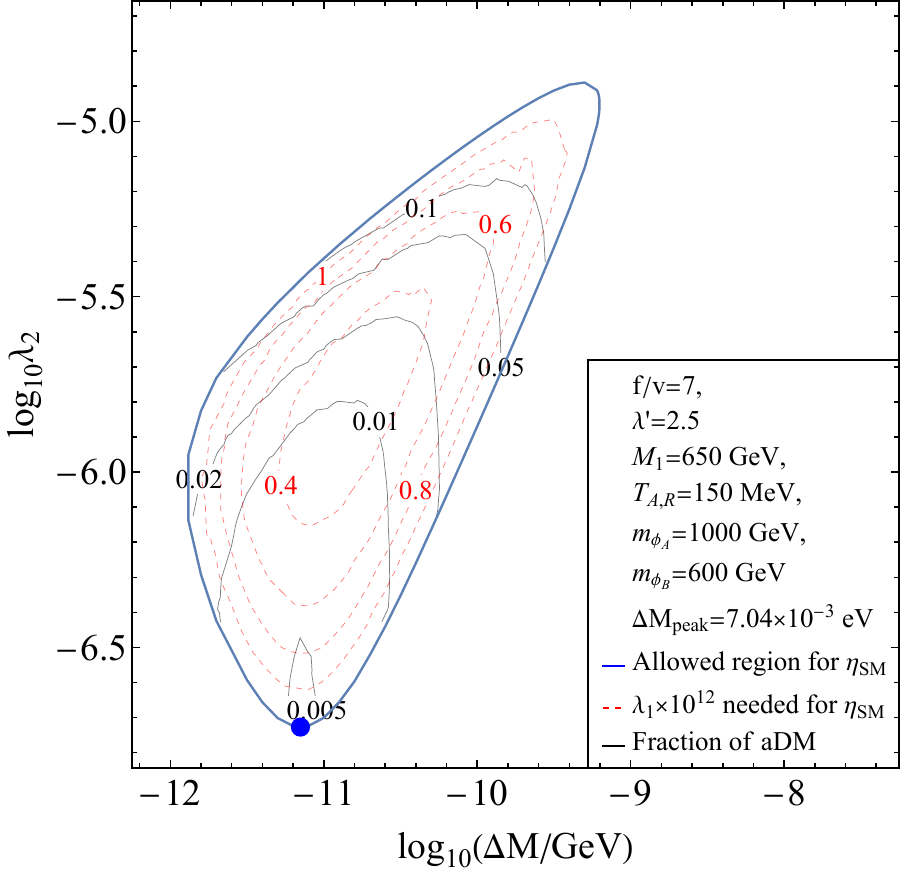}
    \caption{
    Values of the $\lambda_1$ coupling required to generate the visible baryon asymmetry (dashed red contours) in the $(\Delta M,\, \lambda_2)$ plane for the two benchmark scalar masses and top couplings in \cref{e.topbenchmarks}. Outside the blue bounded region, the asymmetric reheating constraint \cref{e.ARHdominant} cannot be satisfied. Black contours show the twin baryon DM fraction $\hat r$. The maximal resonant enhancement occurs for $\Delta M = \Delta M_{peak} \approx \Gamma_{N_2}/2$, which also minimizes the twin baryon asymmetry for the minimal allowed value of $\lambda_2$, and is indicated with a blue dot.
    }
    \label{fig:coup13}
\end{figure}

We therefore focus exclusively on the top-philic $Y_\phi = -1/3$ scenario where the visible baryon asymmetry is generated in 3-body reheaton decays, while the twin bayron asymmetry is generated in 2-body reheaton decays with $m_{\phi_B} < M_{N_1} - m_{b_B}$.\footnote{This means that the generated $\phi_B$ only has 4-body decays, but  in our regime of interest the corresponding lifetime is still prompt and has no effect on the thermal history in the B sector.} 
Generating the observed baryon asymmetry in the A sector fixes a combination of couplings that also control the baryon asymmetry generation in the B sector.
As a consequence, the kinematic suppression in the A sector leads to an overproduction of twin baryons in the B sector. 
This leads us to once more consider the resonant regime 
$M_{N_1} \approx M_{N_2}$.
Since the asymmetry parameters $\epsilon_{CPV}^{A,B} \sim \delta_{A,B}$ depend on different coupling combinations,
\begin{align}
    \label{e.deltatopphilic}
    \delta_A &=  y_1 y_2 \lambda_1^* \lambda_2 {\lambda'}^2, \\
    \delta_B &= (\lambda_1^* \lambda_2)^2,
\end{align} 
the overall resonance enhancement in both sectors  avoids overproduction of aDM when generating the observed SM asymmetry. 
The relevant processes are shown in  \cref{fig:topphilicdiag}.

In this scenario, $\Delta M = M_{N_1} - M_{N_2}$ has to be very close to the maximum enhancement of $\Delta M \approx \Gamma_{N_2}/2$ in order to generate sufficient baryon asymmetry without overproducing twin baryons. This is very consistent with the approximate $U(3)_N^2$ symmetry postulated in \cref{sec:flavor_alignment}. The total width of $N_2$ is given by
\begin{gather}
    \Gamma_{N_2} \simeq \frac{M_2}{16\pi} \left(y_2^2 + 3 \lambda_2^2\left(1-\frac{m_{\phi_B^2}}{m_{N_2}^2}\right)\right) \ ,
\end{gather}
where $y_2$ is fixed by the requirement that $m_{\nu_2} \approx 0.1~\mathrm{eV}$. 
The aDM abundance in the B-sector is then given  by
\begin{align}
    \Omega_{aDM} &= \Omega_b \frac{\epsilon_{\text{CPV,B}}}{\epsilon_{\text{CPV,A}}} \frac{m_{p_{B}}}{m_{p_{A}}} \nonumber \\
    &= \frac{\Omega_b\lambda_1^*\lambda_2}{(y_1y_2)|\lambda'|^2} \frac{ \int d\Pi \Im{\mathcal{A}_0(N\rightarrow b_B+\phi_B)\bar{\mathcal{A}}_1(N\rightarrow b_B+\phi_B})}{\int d\Pi\Im{\mathcal{A}_0(N\rightarrow 2b_A+t_A)\bar{\mathcal{A}}_1(N\rightarrow 2b_A+t_A)}}\frac{m_{p_{B}}}{m_{p_{A}}} .
\end{align}
The dependence of $\eta_{A,B}$ on $\lambda_2$ is therefore non-trivial, since that coupling appears both in the numerator $\delta_{A,B}$ and the denominator due to the resonant enhancement. 
This is shown in \cref{fig:coup13} in the plane of $\Delta M$ and $\lambda_2$ for $f/v = 7, M_{N,1} = 650~\mathrm{GeV}, T_{A,R} = 150~\mathrm{MeV}$, for two benchmark choices of the scalar masses, top couplings and relative phases:
\begin{equation}
    \label{e.topbenchmarks}
    \begin{NiceArray}{l|l|l|l|l|l|l}
    & m_{\phi_A} & m_{\phi_B} & \Delta \xi & \lambda' & \arg(\delta_A) & \arg(\delta_B)\\
    \hline
    \textbf{Benchmark}\ \mathbf{I} & 850~\mathrm{GeV} & 550~\mathrm{GeV} & -0.14 & 1.5  &
    \Block{2-2}<>{\pi/2} & \\
    \cline{1-5}
    \textbf{Benchmark}\ \mathbf{II} & 1000~\mathrm{GeV} & 600~\mathrm{GeV} & -0.22  &2.5  && 
    \end{NiceArray}
\end{equation}
$\Delta \xi = \xi - \xi'$, see \cref{e.mphiBdeltaxi}.
Within the region bounded by the blue contour, the required SM baryon asymmetry can be generated for the values of $\lambda_1$ indicated by the dashed red contours. 
Outside of the blue boundary, the required $\lambda_1$ violates the bound of \cref{e.ARHdominant}. The black contours show the corresponding prediction for the twin baryon DM fraction $\hat r$, spanning from very small values up to overproduction.
The smallest possible $\hat r$ is generated at the very bottom of the bounded region, marked with a blue  dot.
Proton decay constraints on the $\lambda \lambda'$ coupling combinations in \cref{eq:p_decay_loop_23} are easily satisfied in the whole region.
Note that we choose maximal phases for both relevant coupling combinations, corresponding to $\arg(\lambda_1^*\lambda_2) = \pi/4, \arg(\lambda') = \pi/8$ for real $y_{1,2}$. This illustrates the largest possible parameter space to generate the required visible baryon asymmetry. Different untuned choices of these phases would decrease or increase our results for $\Omega_\mathrm{aDM}$ by an $\mathcal{O}(1)$ factor.

\begin{figure}
    \centering
    \includegraphics[width=0.49\linewidth]{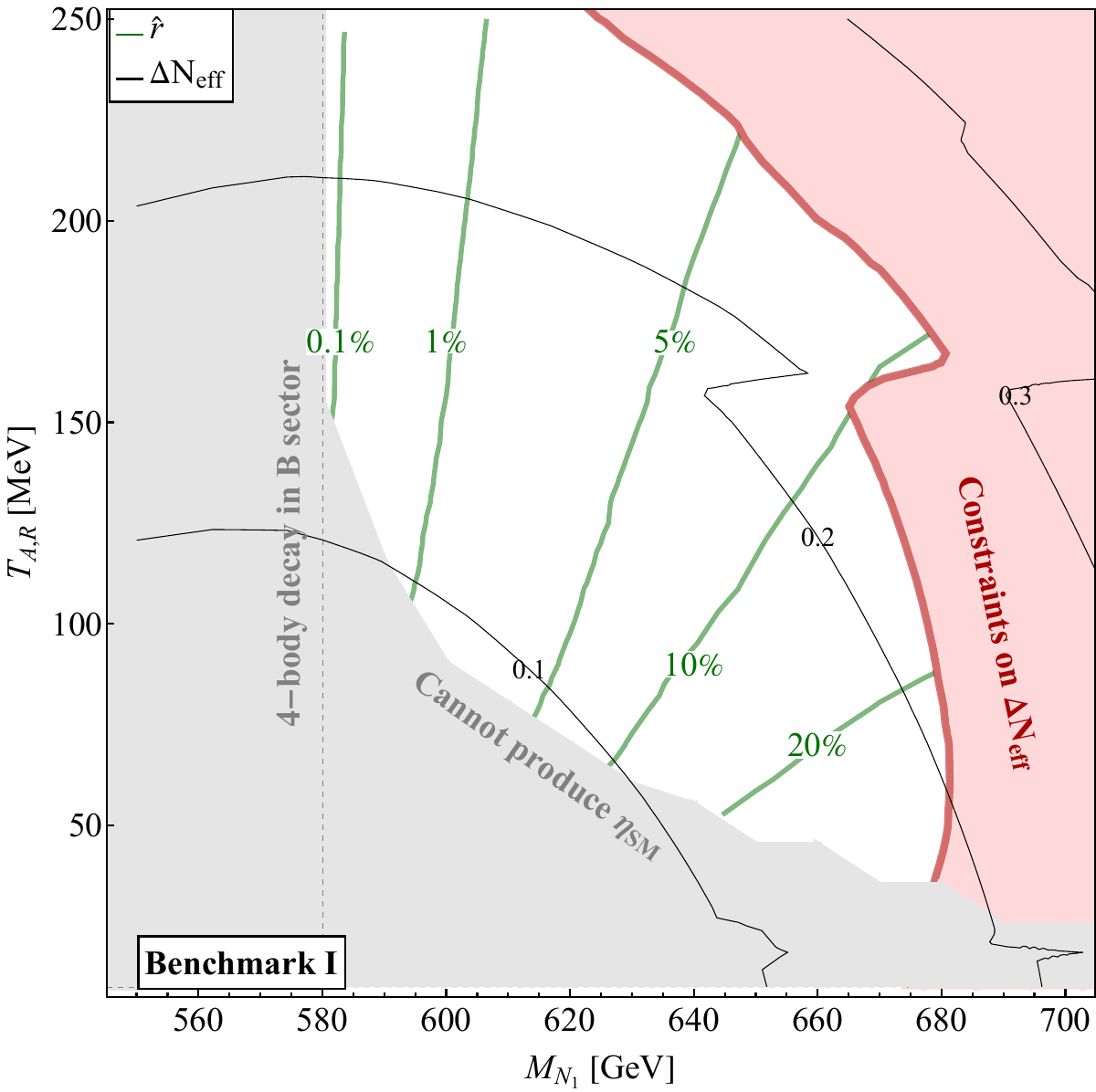}
    \includegraphics[width=0.49\linewidth]{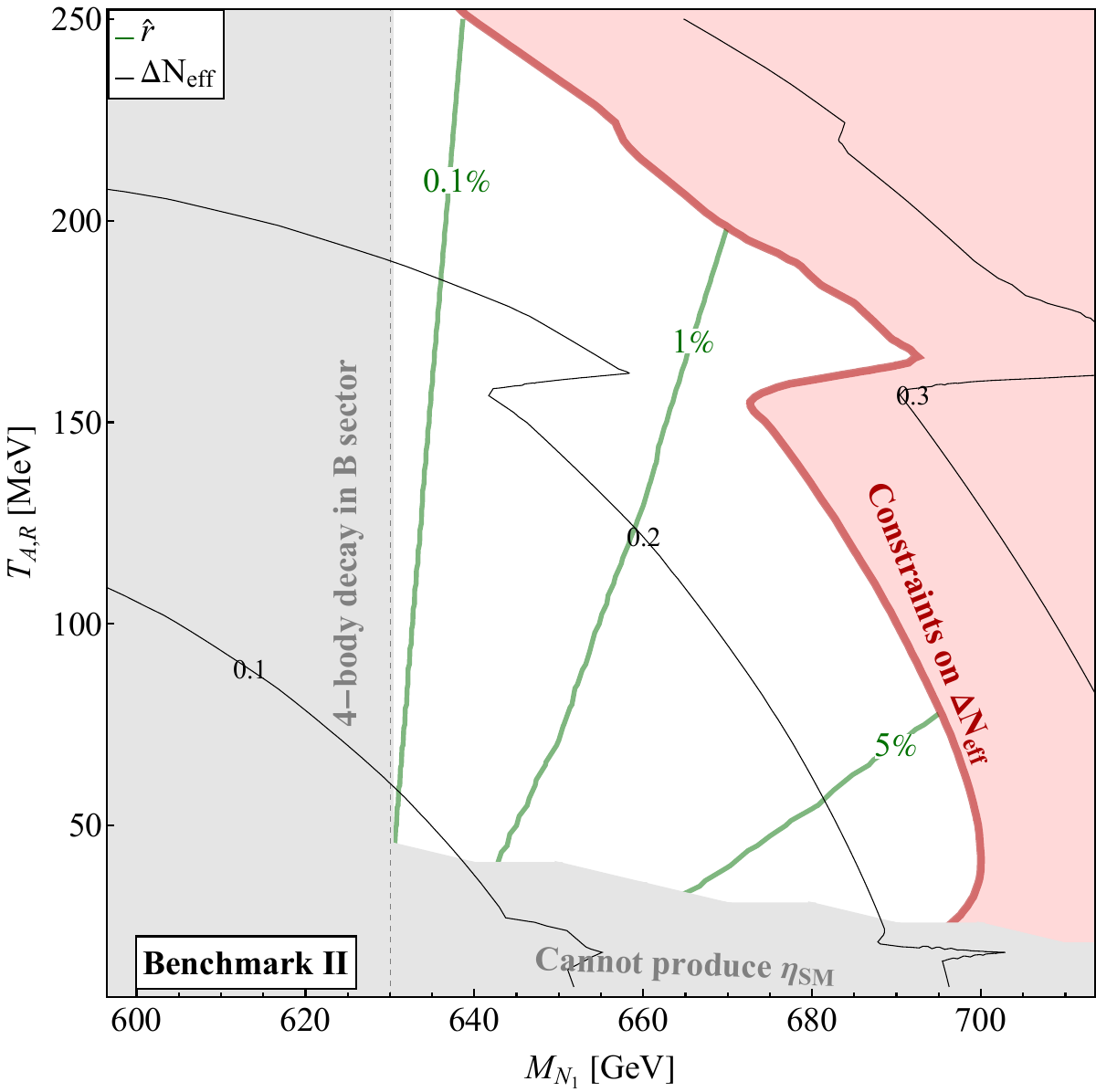}
    \caption{
    Minimum twin baryon DM fraction $\hat r$ (green contours) in the $\nu\phi$MTH model with top-philic couplings and $Y_\phi = -1/3$ as a function of reheaton mass $m_{N_1}$ and visible reheating temperature $T_{A,R}$, for $f/v = 7$.
    On the left, we fix
    $m_{\phi_{A,B}} = 850,~ 550~\mathrm{GeV}, \lambda'= 1.5$;
    on the right, $m_{\phi_{A,B}} = 1000,~600~\mathrm{GeV}, \lambda'= 2.5$, corresponding to the benchmarks in \cref{e.topbenchmarks}.
    We choose $\lambda_2$ and $M_{N_2}$ to maximize the resonant enhancement of $\eta_A$ and thus minimize $\hat r$ (corresponding to the marked blue point in \cref{fig:coup13}). Larger values of $\lambda_2$ within the equivalent of the bounded blue regions of \cref{fig:coup13} could increase $\hat r$ up to $1$ everywhere in this viable parameter space.
    Thin black contours show the $\dneff$ prediction (see \cref{fig:dneff}).
    In the gray area, the bound on $\lambda_1$ implied by \cref{e.ARHdominant} prevents sufficient generation of visible baryon asymmetry.
    In the red shaded area, $\dneff$ and $\hat r$ in combination are excluded by CMB observations~\cite{Bansal:2021dfh}.
    }
    \label{fig:benchmark}
\end{figure}

\begin{figure}
    \centering
    \begin{tabular}{rrr}
    \includegraphics[width=0.9\linewidth]{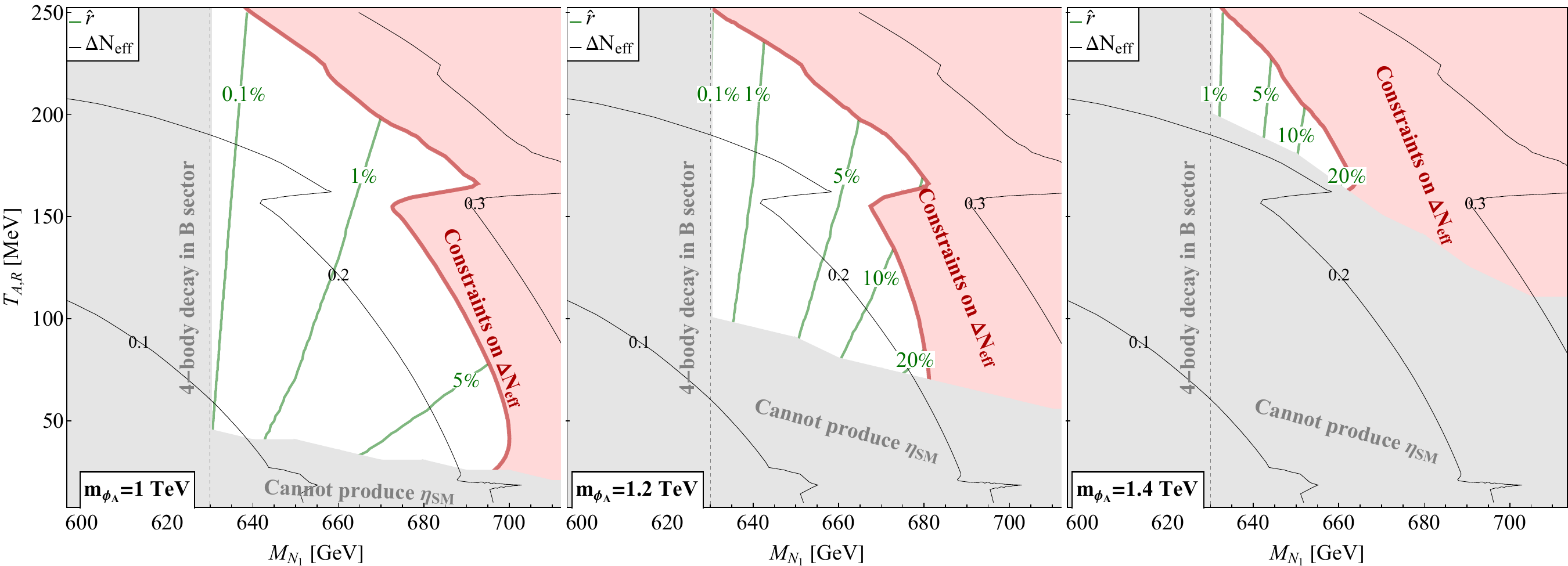}
    \\
    \includegraphics[width=0.9\linewidth]{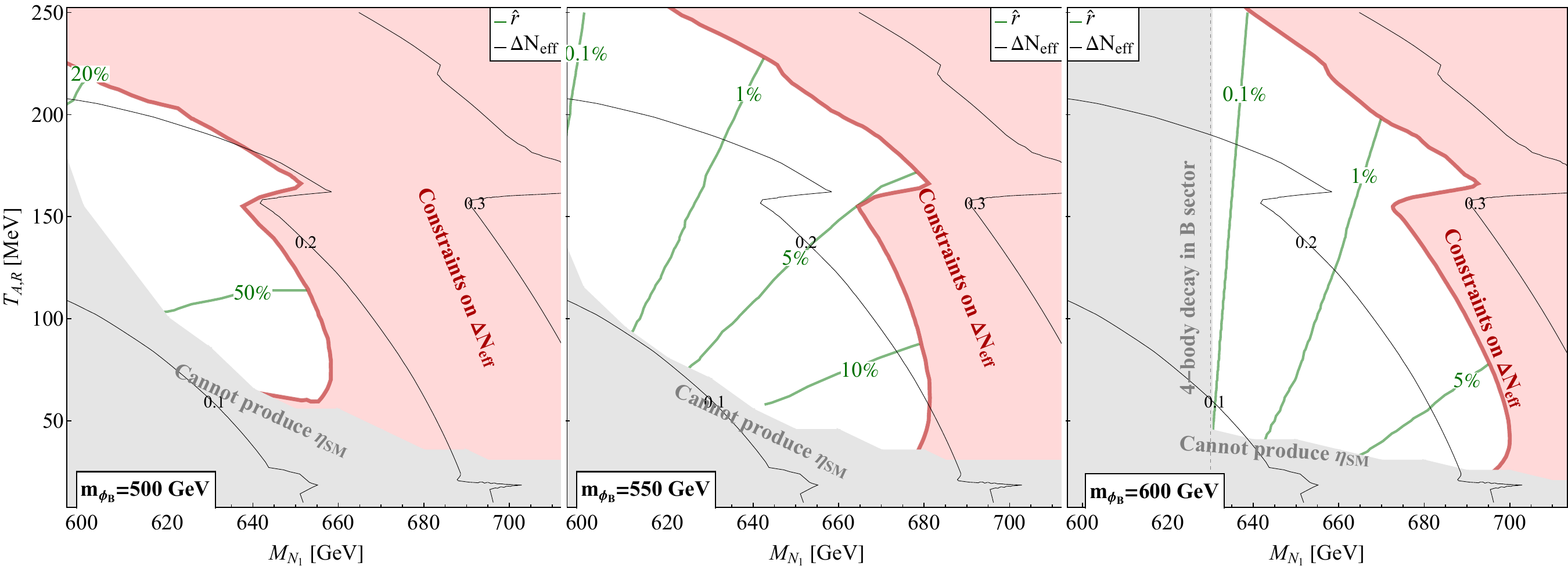}
    \\    
    \includegraphics[width=0.9\linewidth]{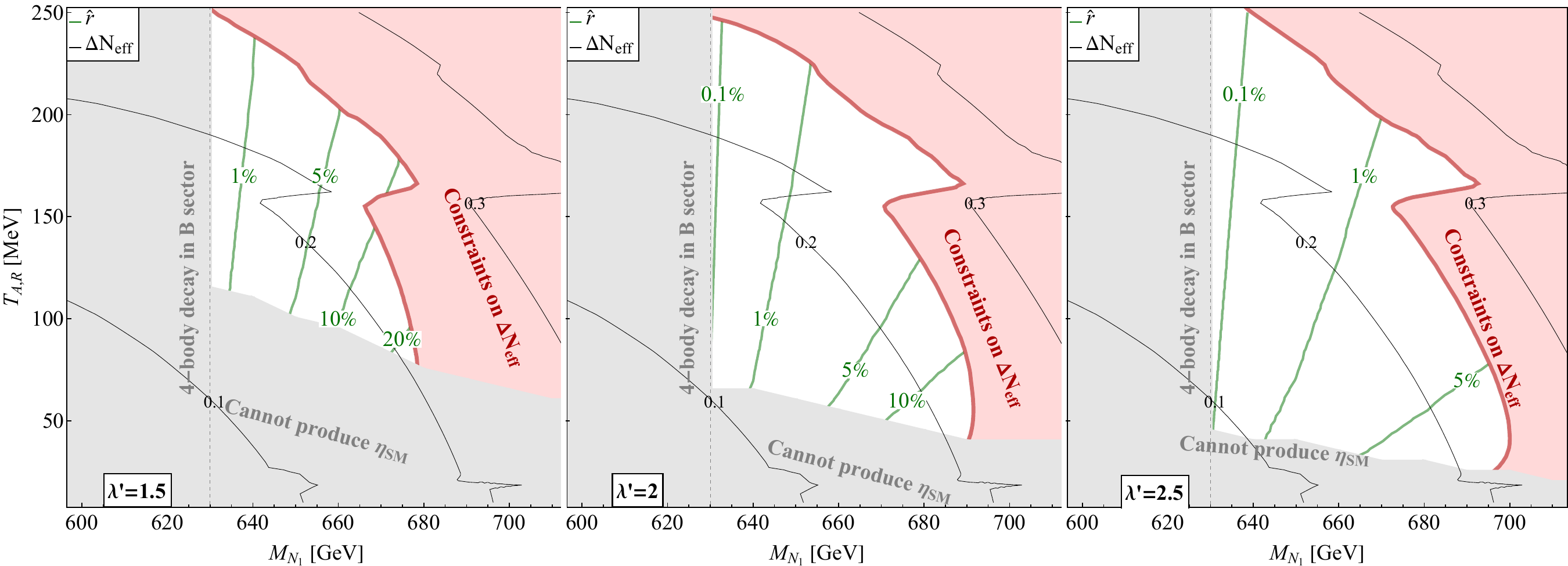}
    \\
     \includegraphics[width=0.9\linewidth]{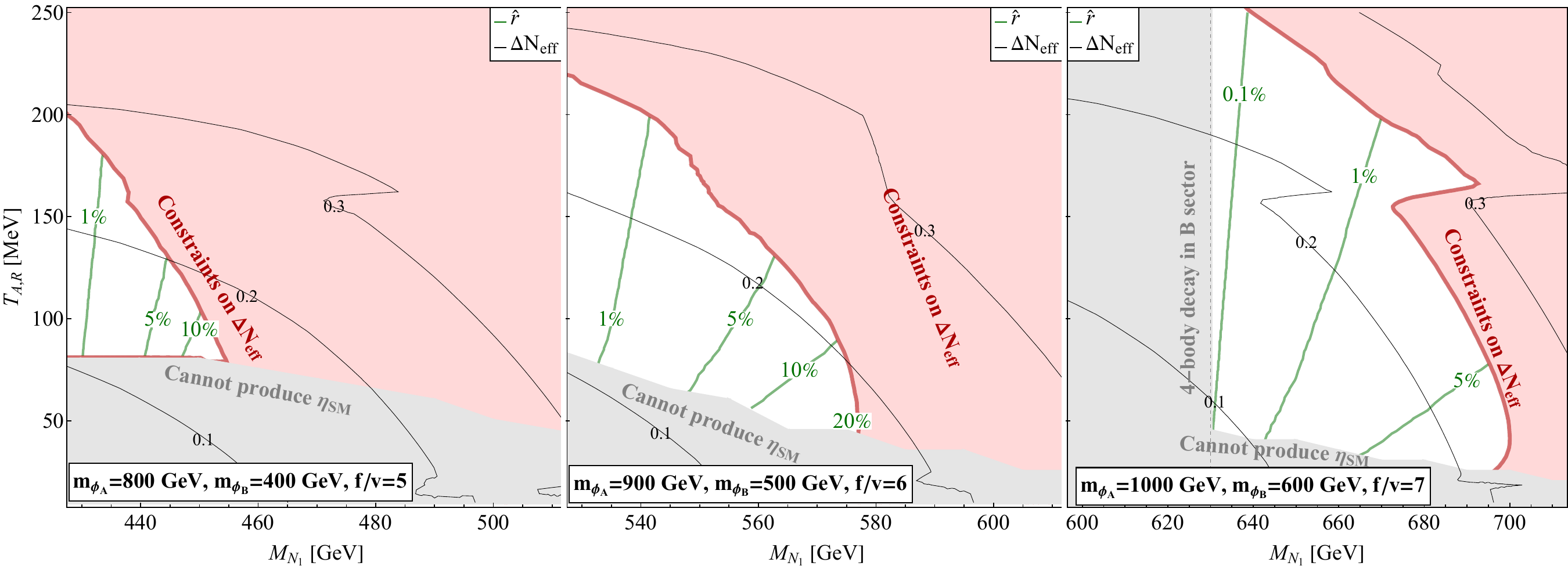}
     \end{tabular}
    \caption{
    Same as \cref{fig:benchmark} (left) for single-parameter variations around  the second benchmark in \cref{e.topbenchmarks}.
    We demonstrate the effect of changing $m_{\phi_A}$ (first row), $m_{\phi_B}$ (second row), $\lambda'$ (third row) and $f/v$ (fourth row, in this case other parameters are also shifted away from the benchmark values to better capture the available parameter space).
    }
    \label{fig:final}
\end{figure}

In \cref{fig:benchmark}, we show the minimum twin baryon asymmetry (for maximal phases) generated for the two benchmark choices of \cref{e.LY23} as a function of reheaton mass and reheating temperature. 
Clearly, the top-philic scenario can generate a wide range of twin baryon asymmetries while satisfying current cosmological constraints (including the dark acoustic oscillations of the twin baryons~\cite{Bansal:2021dfh}). 
The reheating temperature is fairly unconstrained by the additional requirements of baryogenesis, 
except for the fact that very low $T_{A,R}$ near the MeV scale correspond to such small $y_1$ couplings that it becomes difficult to generate the observed visible baryon asymmetry. This also means that the twin baryons undergo  BBN with thermal initial neutron-proton abundances, see \cref{fig:dneff}.
The reheaton and scalar masses have to obey 
\begin{equation}
    \label{e.mN1topphilic}
    m_{\phi_B} + m_{b_B} < M_{N_1} < m_{\phi_A} + m_{t_A} + 2 m_{b_A} 
\end{equation}
in order for the decay to be 3- and 2-body in the visible and hidden sectors respectively.

In  \cref{fig:final} we show the effect of changing the scalar masses $m_{\phi_{A,B}}$, scalar top couplings $\lambda'$ and $f/v$  around the second benchmark in \cref{e.topbenchmarks}. This demonstrates that the $\nu\phi$MTH model has a large viable parameter space with colored scalar masses above the TeV scale, and astrophysically and cosmologically interesting twin baryon abundances.

In the scenarios discussed above, $\lambda'$ is taken to be $\mathcal{O}(1)$. So far, we have assumed that $\lambda'' = 0$, but now let us consider what happens for nonzero values. 
It is straightforward to apply all of the above calculations to this case, as long as $\lambda'$ dominates the three-body decay in the A-sector.
In that case, \cref{fig:coup13} and the proton decay bound \cref{eq:p_decay_loop_23} imply $\lambda'' \lesssim 10^{-2}$. Satisfying this bound also ensures that no additional significant decay channels complicate our above analysis. 
We therefore require a modest hierarchy $\lambda'' < \lambda'$. One way this could arise is if  $\phi_{A,B}$ were identified with squarks in a supersymmetric Twin Higgs UV completion, as discussed near the end of \cref{sec:model_defn}. However, the significant mass difference between $m_{\phi_A}$ and $m_{\phi_B}$ to avoid light new colored states in the visible sector requires quartic couplings of order $\Delta \xi \sim \mathcal{O}(0.1)$. 
This is much larger than the $\sim y_b^2$ value predicted if we identify $\phi_{A,B}$ with (twin) sbottoms. Therefore, a consistent supersymmetric interpretation of a natural suppression of $\lambda''$ relative to $\lambda'$ would require identification of $\phi_{A,B}$ with an exotic down-type squark in the UV completion. 

Since the $\lambda'$ coupling has to be sizeable, we comment on its RG evolution. The color-antisymmetric interaction vertex means that $\lambda'$ does not contribute to its own running at 1-loop order. However, the gluon exchange vertex correction does contribute, with $(-\frac{1}{2})\times$ the color factor of the corresponding term in the top Yukawa RG. For both of our benchmarks, this suggests that the $\lambda'$ coupling remains perturbative well beyond the PeV scale. This does not impose any new restrictions on the Twin Higgs UV completion scale, which can be at most $\sim 5 - 10$~TeV.

Finally, we comment on the required quark alignment of the top-philic scenario. Depending on the dynamics  underpinning flavor, there could be rotations between the right-handed quark mass and flavor bases. Proton decay constraints restrict these  rotation angles to be below $\sim 10^{-5}$, see \cref{eq:flavormixing2}.

We conclude this section by presenting \cref{t.mainsummary}, which summarizes all the various requirements on the $\nu\phi$MTH parameters in the top-philic $Y_\phi = -1/3$ scenario.

\begin{table}
    \centering
    \vspace*{-10mm}
    \hspace*{-4mm}
    \begin{tabular}{|m{0.14\textwidth}|m{0.85\textwidth}|}
    \hline
    \multicolumn{2}{|c|}{}
    \\ [-3mm]
    \multicolumn{2}{|c|}{
    $\mathcal{L} =   \mathcal{L}_{A} +  \mathcal{L}_{B} + \mathcal{L}_{\nu\textrm{MTH}} + \mathcal{L}_{\phi,\text{int}}- V$
    }
    \\
    \multicolumn{2}{|c|}{}
    \\ [-3mm]
    \multicolumn{2}{|c|}{
    $\mathcal{L}_{\nu MTH}  = - (y_i H_A \bar{L}_{A,i} N_{A,i} + \frac{1}{2}M_{N_i}\bar{N}_{A,i} N_{A,i}^c) - (A \to B) - M_{AB,ij}\bar{N}_{A,i} N^c_{B,j}$ 
    }
    \\
    \multicolumn{2}{|c|}{}
    \\ [-3mm]
    \multicolumn{2}{|c|}{
    $\mathcal{L}_{\phi,\text{int}} = -  \lambda_{i} \phi^{}_A  b_{A}^c \bar{N}_{i,A} - \lambda' \phi_A^\dagger \bar{t}^{}_{A} b_{A}^c +  \lambda'' \phi_A^\dagger \bar{Q}^{}_{3,A} Q_{3,A}^c  + (A \to B)$
    }
    \\
    \multicolumn{2}{|c|}{}
    \\ [-3mm]
    \hline
    \textbf{Parameter} & \textbf{Requirement and possible theory interpretation} \\
    \hline \hline
    $m_{N_1}, y_1$ & 
    Baryogenesis and asymmetric reheating require $m_h < m_{N_1}$ and $m_{N_1} \lesssim m_{W_B}$. Therefore $125~\mathrm{GeV} < m_{N_1} \lesssim 700~\mathrm{GeV}$
    for natural values of $f/v \lesssim 7$. \phantom{blabla}
    Sufficient lifetime of $N_1$ then requires $y_1 \sim 10^{-12} - 10^{-10}$, corresponding to a tiny lightest active neutrino mass $m_{\nu_1} \sim (10^{-14} - 10^{-10})~\mathrm{eV}$.
    \\ 
    \hline
    $m_{N_2}, y_2$
    &
    $\nu$ oscillation data requires $m_{\nu_{2}} \sim 0.1~\mathrm{eV}$, setting
    $ y_{2} \sim 10^{-6} \sqrt{M_{N_{2}}/(500~\mathrm{GeV})}$. 
    \\
    \hline
    Lepton Flavor Alignment
    &
    The single dominant reheaton $N_1$ with  $m_{N_1} \gg~\mathrm{GeV}$ and $y_1 \ll y_2,y_3$ requires that $y$-hierarchies and mixings are generated separately, and that $N$-masses are approximately universal, see \cref{sec:flavor_alignment}.
    \\ \hline
    $\Delta M_{AB}$
    &
    Cosmological bounds on active-sterile neutrino mixing require $M_{AB}/M_N \lesssim 10^{-3} - 10^{-2}$~\cite{Hannestad:2012ky} $\Rightarrow$ $\mathbb{Z}_2$ even/odd $N_{i\pm}$ mass eigenstates are near-degenerate.
    \\ 
    \hline
    $m_{\phi_A}, \lambda'$
    & 
    $m_{\phi_A} \gtrsim~\mathrm{TeV}$
    without violating proton decay constraints if the visible baryon asymmetry is generated via three-body decays $N_1 \to \phi_A^*$, requiring $\lambda' \sim \mathcal{O}(1)$.
    \\
    \hline
    $m_{\phi_B}$ 
    &
    Generating twin baryon abundance requires $N_1 \to \phi_B$ decays in the twin sector to be two-body: $
        m_{\phi_B} + m_{b_B} < m_{N_1} < m_{\phi_A} + m_{t_A} + 2 m_{b_A} $.
    \\
    \hline
    $m_{N_1} - m_{N_2}$ 
    &
    Avoiding overproduction of twin baryons while generating sufficient visible baryons requires resonant enhancement $m_{N_1} \approx m_{N_2}$, with the mass difference being of similar scale as the radiative contribution $\Delta M \sim \Gamma_{N_2}$ (since $y_2 \gg y_1$).
    Together with small $N_i$ mixing requirement 
    to protect $\lambda_i, y_i$ hierarchies, this 
    suggests the approximate $U(3)_N^2$ flavor symmetry postulated in \cref{sec:flavor_alignment}.
    \\
    \hline
    $\lambda_1, \lambda_2$
    &
    Preserving asymmetric reheating, see \cref{e.ARHdominant}, requires $\lambda_1/y_1 \lesssim 0.1$. Generating the observed SM baryon abundance then selects $\lambda_2 \sim y_2$.
    \\
    \hline
    $\lambda''$
    &
    $\lambda'' \lesssim 10^{-2}$ from proton decay constraints \cref{eq:p_decay_loop_23}.
    The suppression relative to $\lambda'$ could be a consequence of $\phi_{A,B}$ being exotic down-type squarks within a supersymmetric Twin Higgs UV completion, with Higgs coupling larger than $y_b^2$ to generate the required mass difference between $\phi_A$ and $\phi_B$. 
    \\
    \hline
    Quark Flavor Alignment
    &
    The UV dynamics generating flavor have to align the right-handed quark mass and flavor bases (in which  $\lambda, \lambda', \lambda''$ couplings are top-philic) at a level better than  $\sim 10^{-5} - 10^{-4}$ to satisfy proton decay constraints, see \cref{eq:flavormixing2}. 
    \\ \hline
    $f/v$ & 
    $f/v \gtrsim 5$ to jointly satisfy viable SM baryogenesis, small $\dneff$ and $m_{\phi_A} \gtrsim~\mathrm{TeV}$ for perturbative $\lambda'$ due to $\epsilon_{CPV}^A \propto {\lambda'}^2 (m_{N_1}/m_{\phi_A})^4$.
    \\ \hline
    \end{tabular}
    \caption{
    Summary of the top-philic $\nu\phi$MTH model with (twin-)hypercharge $Y_\phi = -1/3$ for the (twin-)colored scalars $\phi_{A,B}$.
    The Lagrangian is shown at the top, the requirements on its parameters in order to
    satisfy cosmological constraints via asymmetric reheating and generate SM and twin baryon number are given in the table.
    The Lagrangian is written in the quark mass basis and in the $\nu_L, N$ basis most aligned with their mass bases before active-sterile mixing.
    $y$-couplings are therefore diagonal.
    The scalar potential $V$ is given in \cref{e.fullpotential}.
    $N_1$ is the reheaton decaying dominantly into the $A$-sector after a period of matter domination, with a small fraction of its decays via $\phi_{A,B}^{(*)}$ and a resonant $N_2$ generating visible and twin baryon asymmetries. $N_3$ has properties similar to $N_2$ but without the requirement of being very close in mass to $N_1$, and we assume the effect of $N_2$ in $N_1$-decay dominates baryogenesis. 
    }
    \label{t.mainsummary}
\end{table}

\section{Conclusions}
\label{sec:conclusion}

Theories of Neutral Naturalness like the Twin Higgs~\cite{Chacko:2005pe, Chacko:2016hvu, Craig:2016lyx}, can solve the Little Hierarchy problem without introducing colored top partners that have not been observed at the LHC. The twin mechanism relies on a discrete $\mathbb{Z}_2$ symmetry that relates the SM to a hidden sector copy containing versions of the same particles and forces. 
The cosmology of this scenario is nontrivial, requiring either hard $\mathbb{Z}_2$ breakings to eliminate excess dark radiation, or some kind of asymmetric reheating to lower the temperature of the hidden sector relative to the visible sector.

In this work we present the $\nu\phi$MTH model, for the first time realizing a Mirror Twin Higgs model that produces visible and twin baryon asymmetries in the decay of the same reheaton that asymmetrically reheats the two sectors to satisfy cosmological constraints, with only the minimal soft $\mathbb{Z}_2$ breaking required by Higgs measurements.
Our model is based on the 3-generation $\nu$MTH model~\cite{Chacko:2016hvu}, see \cref{eqn:nuMTH}, where all active neutrino masses are generated by a type-1 see-saw  and one of the right-handed neutrinos acts as the reheaton. 
New \mbox{(twin-)}colored scalars $\phi_{A,B}$ coupling to (twin-)quarks and right-handed neutrinos allow a fraction of reheaton decays to generate (twin) baryons. 

We conduct a detailed study of various possible implementations of this mechanism, with different (twin-) hypercharges and quark coupling structures of the new (twin-) colored scalars.
Satisfying proton decay constraints, and avoiding the mass of the new colored scalar $\phi_A$ being below the TeV scale to preserve the collider phenomenology of neutral naturalness,  
uniquely selects a $\phi_{A,B}$ with (twin-) hypercharge $Y_\phi = -1/3$ and \emph{top-philic} Yukawa couplings $\lambda, \lambda', \lambda''$ to bottoms and right-handed neutrinos, right-handed top and bottom quarks, and left-handed quarks, see \cref{eq:L_phi_13}. 

This viable model is analyzed in detail in \cref{s.topphilic} and summarized in \cref{t.mainsummary}. 
It can generate the observed SM baryon asymmetry, while also generating a wide range of twin baryon asymmetries, yielding atomic dark matter fractions $\hat r = \Omega_{aDM}/\Omega_{DM}$ from negligible to unity. 
Baryogenesis in both sectors requires  $\dneff \sim 0.1$, which may be detected in future observations. This mostly corresponds to reheating temperatures where twin BBN proceeds normally, resulting in the expected twin helium mass fraction $\sim 0.6-0.7$ of aDM~\cite{Bansal:2021dfh}.
The twin Higgs vev ratio is favored to lie above $f/v \gtrsim 5$ for viable baryogenesis and sufficiently small $\dneff$ with $m_{\phi_A} \gtrsim~\mathrm{TeV}$ and perturbative $\lambda'$ couplings, which is reasonably natural but may motivate constructions like~\cite{Beauchesne:2015lva, Harnik:2016koz, Csaki:2019qgb, Durieux:2022sgm} that generate such hierarchies without tuning.

Our $\nu\phi$MTH model has a variety of distinctive theoretical and phenomenological features that arise from the combined requirements of asymmetric reheating and the generation of baryon asymmetries in both sectors without hard $\mathbb{Z}_2$ breakings, which are also summarized in \cref{t.mainsummary}. The coupling hierarchies can be succinctly presented as
\begin{equation*}
    \lambda_1 < y_1 \ll y_2 \sim \lambda_2 \ll \lambda'' \ll \lambda' \sim \mathcal{O}(1) ,
\end{equation*}
where the hierarchies amongst the tiny $y, \lambda$ Yukawa couplings, as well as the top-philic alignment and generation of anarchic $\nu_L$ mixing angles, can have their origin in some underlying flavor dynamics or symmetries~\cite{Knapen:2015hia} as discussed in \cref{sec:flavor_alignment}. The same could be true for the $\lambda^{\prime(\prime)}$ hierarchy, but this can also be a consequence of a supersymmetric UV completion. 
Cosmological constraints and generation of the required visible baryon asymmetry may still be satisfied when departing from this regime, but this would result in negligible twin baryon density. 

As we discuss in \cref{s.genuniversal}, the scenario with generation-universal quark couplings $\lambda, \lambda^{\prime(\prime)}$ can also yield a viable cosmology with the correct SM baryon asymmetry, but this requires $m_{\phi_A} < M_{N_1} \lesssim m_{W_B}$, resulting in a new colored scalar below the TeV scale. 
This can only be consistent with current bounds if  $m_{\phi_A} \approx 550~\mathrm{GeV}$ and this colored scalar is in fact responsible for the 2-3 sigma excess in the recent CMS search~\cite{CMS:2022usq}. Apart from the fact that this runs somewhat counter to the phenomenological motivation for neutral naturalness, this scenario may be excluded soon in updated analyses.

Our work demonstrates that unifying the mechanisms of asymmetric reheating and baryogenesis in MTH models is surprisingly involved. It would be interesting to investigate other possible asymmetric reheating mechanisms beyond the decay of right-handed neutrinos, to understand whether this is generic or could be accommodated more easily in a different setup.
A priori, other very low-scale baryogenesis mechanisms like the one proposed in~\cite{Elor:2018twp,Alonso-Alvarez:2019fym,Alonso-Alvarez:2021qfd} can be compatible with (and in fact benefit from) the asymmetric reheating requirement, though the additional twin sector dynamics must be carefully taken into account.
Depending on the outcome of these considerations, it could also be interesting to understand the general requirements that baryogenesis places on the UV completion of the Twin Higgs. We leave this for future investigation.

In general terms, our top-philic $\nu\phi$MTH model predicts that some fraction of dark matter is  twin baryons, with a wide range of possible abundances that are highly astrophysically and cosmologically relevant. 
This realizes a particular atomic dark matter scenario with dark electron mass $(f/v)\times m_{e}$, SM-QED-like coupling, and dark proton masses 30 - 70\% heavier than SM protons~\cite{Chacko:2018vss}, with the additional feature of dark nuclear interactions and twin BBN that converts the majority of twin nucleons into twin helium~\cite{Bansal:2021dfh}.
The wide range of possible aDM fractions motivates the search for signatures of atomic dark matter on cosmological~\cite{Cyr-Racine:2013fsa,
Bansal:2022qbi,
Bansal:2021dfh,
Zu:2023rmc} and galactic~\cite{Fan:2013yva,Fan:2013tia,McCullough:2013jma,Randall:2014kta,Schutz:2017tfp, Buch:2018qdr, Ghalsasi:2017jna, Ryan:2021dis, Gurian:2021qhk, Ryan:2021tgw, Foot:2013lxa,  Foot:2014uba, Foot:2015mqa,Chashchina:2016wle, Foot:2017dgx, Foot:2018dhy, Foot:2016wvj, Foot:2013vna,Roy:2023zar, Gemmell:2023trd}
scales as well as direct detection experiments~\cite{Chacko:2021vin}, while the existence of dark nuclear interactions motivates specific searches for dark-fusion-supported mirror stars~\cite{Curtin:2019ngc, Curtin:2019lhm, Howe:2021neq, Winch:2020cju}, mirror neutron stars~\cite{Hippert:2021fch, Hippert:2022snq}, and mirror white dwarfs~\cite{Ryan:2022hku}, which could be detected via their (faint) electromagnetic, gravitational lensing or gravitational wave signals. The solution to the electroweak hierarchy problem may therefore be discovered first in the forthcoming treasure trove of detailed astrophysical data, rather than direct searches at high energy colliders.

\acknowledgments

We thank Wolfgang Altmannshofer, Jared Barron, Zackaria Chacko, Micah Mellors and Yuhsin Tsai for helpful conversations. 
The work of GAA, DC, AR and ZY was supported in
part by Discovery Grants from the Natural Sciences
and Engineering Research Council of Canada, the
Canada Research Chair program,  the Alfred P. Sloan Foundation,
the Ontario Early Researcher Award, and the University of Toronto McLean Award. The work of AR was also supported by  the University of Toronto Connaught International Scholarship.

\appendix

\bibliographystyle{JHEP}
\bibliography{biblio.bib}

\end{document}